\documentclass{amos}
\usepackage{subfig}
\usepackage{amsmath}
\usepackage{amsfonts}
\usepackage{xcolor}
\usepackage{multirow}
\usepackage{url}
\usepackage{floatrow}
\usepackage[export]{adjustbox}

\title{\TitleFont High Frequency, High Accuracy Pointing onboard Nanosats using Neuromorphic Event Sensing and Piezoelectric Actuation}
\author{
Yasir Latif$^1$, 
Peter Anastasiou$^2$, 
Yonhon Ng$^3$, 
Zebb Prime$^2$,
Tien-Fu Lu$^1$,
Matthew Tetlow$^2$,
Robert Mahony$^3$,
and Tat-Jun Chin$^1$, \\ 
$^1$The University of Adelaide, 
$^2$Inovor Technologies,
$^3$Australian National University}

\date{}

\begin{document} 
\maketitle

\begin{abstract}
	\normalsize
	As satellites become smaller, the ability to maintain stable pointing decreases as external forces acting on the satellite come into play. At the same time, reaction wheels used in the attitude determination and control system (ADCS) introduce high frequency jitter which can disrupt pointing stability. For space domain awareness (SDA) tasks that track objects tens of thousands of kilometres away, the pointing accuracy offered by current nanosats, typically in the range of 10 to 100 arcseconds, is not sufficient. In this work, we develop a novel payload that utilises a neuromorphic event sensor -- for high frequency and highly accurate relative attitude estimation -- paired in a closed loop with a piezoelectric stage -- for active attitude corrections -- to provide highly stable sensor-specific pointing. Event sensors are especially suited for space applications due to their desirable characteristics of low power consumption, asynchronous operation, and high dynamic range. We use the event sensor to first estimate a reference background star field from which instantaneous relative attitude is estimated at high frequency. The piezoelectric stage works in a closed control loop with the event sensor to perform attitude corrections based on the discrepancy between the current and desired attitude. Results in a controlled setting show that we can achieve a pointing accuracy in the range of 1-5 arcseconds using our novel payload at an operating frequency of up to 50Hz using a prototype built from commercial-off-the-shelf components. Additional results and video demos can be found in the accompanying repository\footnote{ \url{https://www.yasirlatif.info/ultrafinestabilisation}} online.

\end{abstract}

%%%%%%%%%%%%%%%%%%%%%%%%%%%%%%
\section{Introduction}\label{sec:introduction}

\begin{figure}[!h]
    \centering
    \includegraphics[width=0.35\textwidth]{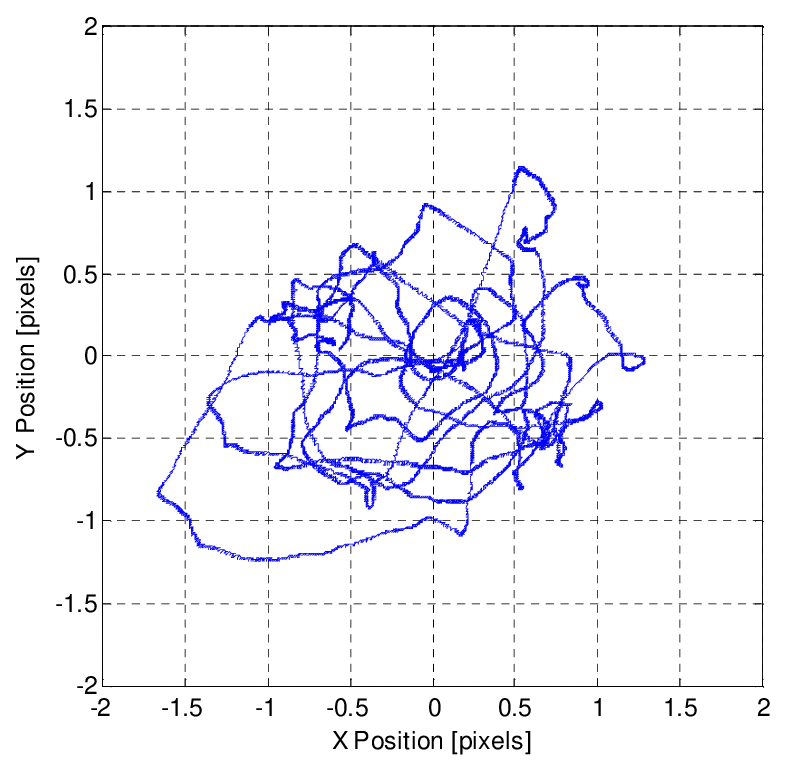}
    \includegraphics[width=0.35\textwidth]{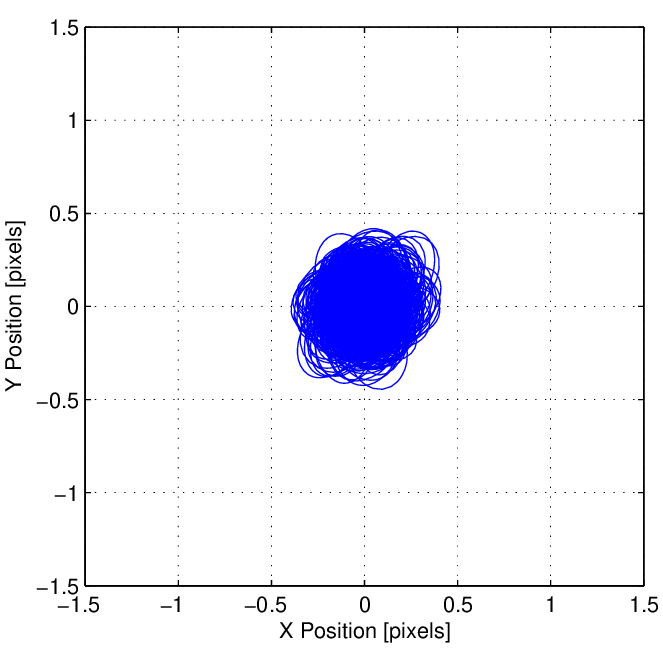}
    \caption{
    \textbf{Left}: Scattering of photons on the image plane due to CubeSat jitter. 
    \textbf{Right}: Improved pointing accuracy enables the light to be focused on a smaller image region (e.g., within a single pixel). Figure adapted from \cite{pong2010achieving} 
    %\textbf{(right)} High-speed event stream (in $\mu s$ time resolution) from observing a star field under jitter.
    }
    \label{fig:motivation}
\end{figure}

Many commercially important space-based applications require precise control of spacecraft attitude. This includes applications that require consistent pointing of onboard sensors towards Resident Space Objects (RSOs), regions in a distant orbit or on the Earth’s surface. 
%Similarly, establishing optical communication links in space also requires attitude control with a high degree of stability.
%
Precise pointing plays a critical role when the imaging sensor needs to be exposed for a long duration -- ten of seconds to minutes -- to capture enough reflected light from a distant RSO. Any pointing errors in the spacecraft during this exposure time cause the precious photons from the RSO to be spread over multiple pixels (Fig. \ref{fig:motivation}~(left)), reducing the Signal to Noise Ratio (SNR) as well as the probability of detection. Precise stabilisation can help focus the light on a much smaller region within the imaging sensor (Fig.~\ref{fig:motivation}~(right)). Such stabilisation becomes mission critical when the main mission sensor experiences high frequency perturbations that can not be characterised by the sensor due to its low sampling rate. Therefore, there is a need for a a higher sampling rate sensor that can characterise such perturbations. 

\begin{figure*}
    \centering
    \subfloat[][Space-time event volume]{
    \includegraphics[width=0.49\textwidth]{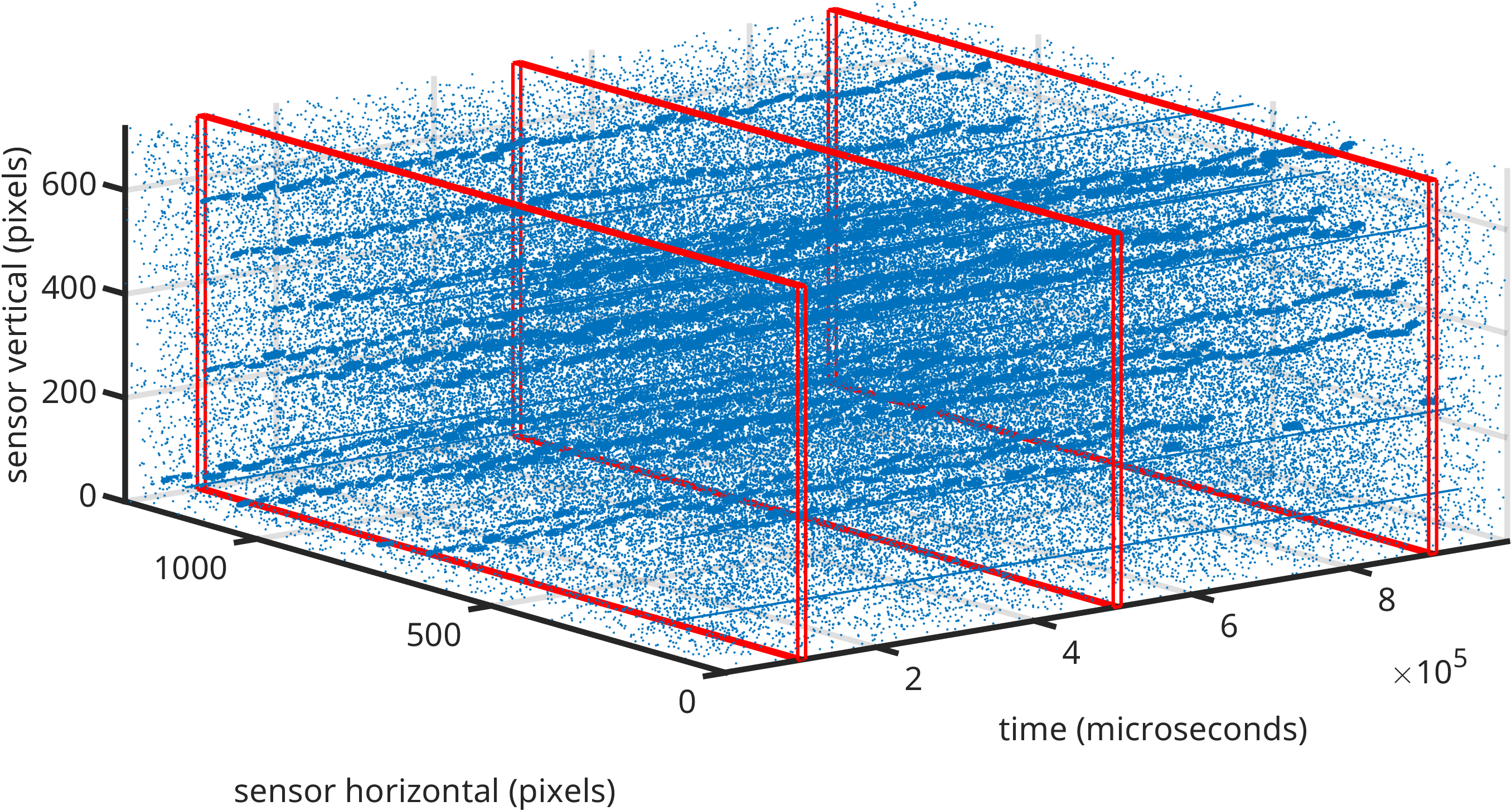}
    \label{fig:xyt-volume}}
    \subfloat[][Events horizontal location (x) against time]{
    \includegraphics[width=0.49\textwidth]{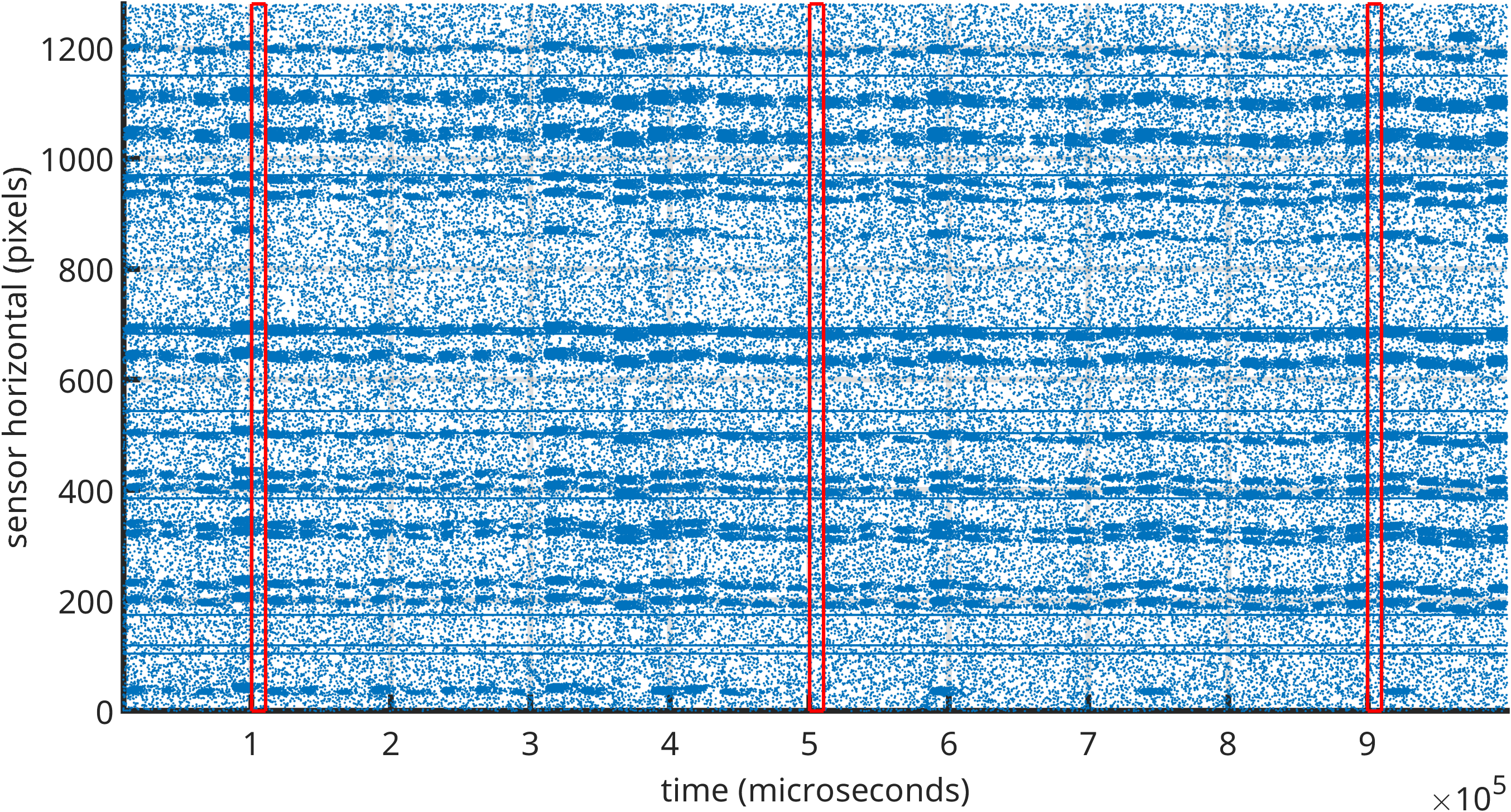}
    \label{fig:xt-slice}}
    \\
    \includegraphics[width=0.49\textwidth]{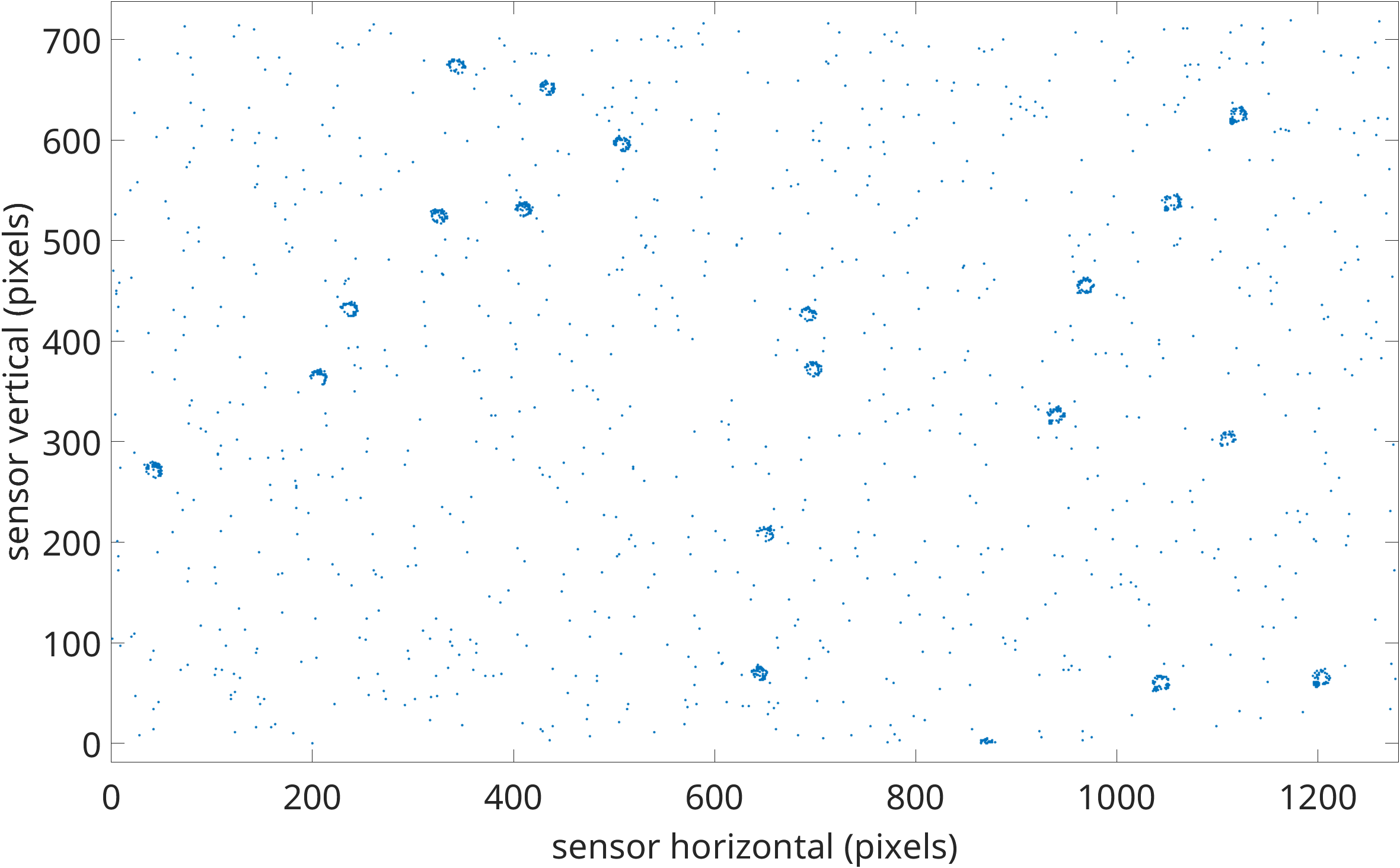}
    \includegraphics[width=0.49\textwidth]{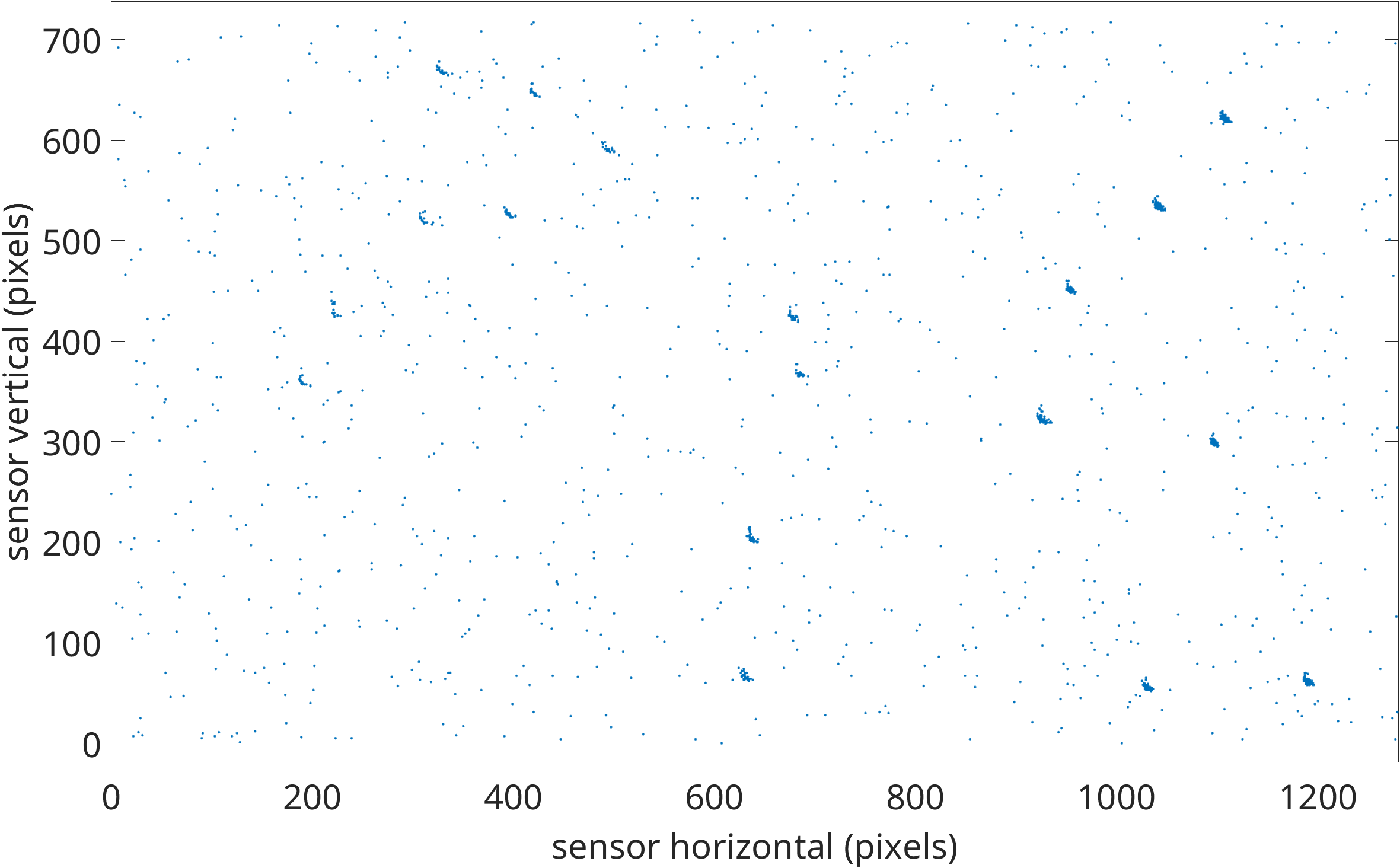}\\
    \subfloat[][(1-3) Events sensed in 10ms duration marked in (b). (4) 1 second long exposure]{
    \includegraphics[width=0.49\textwidth]{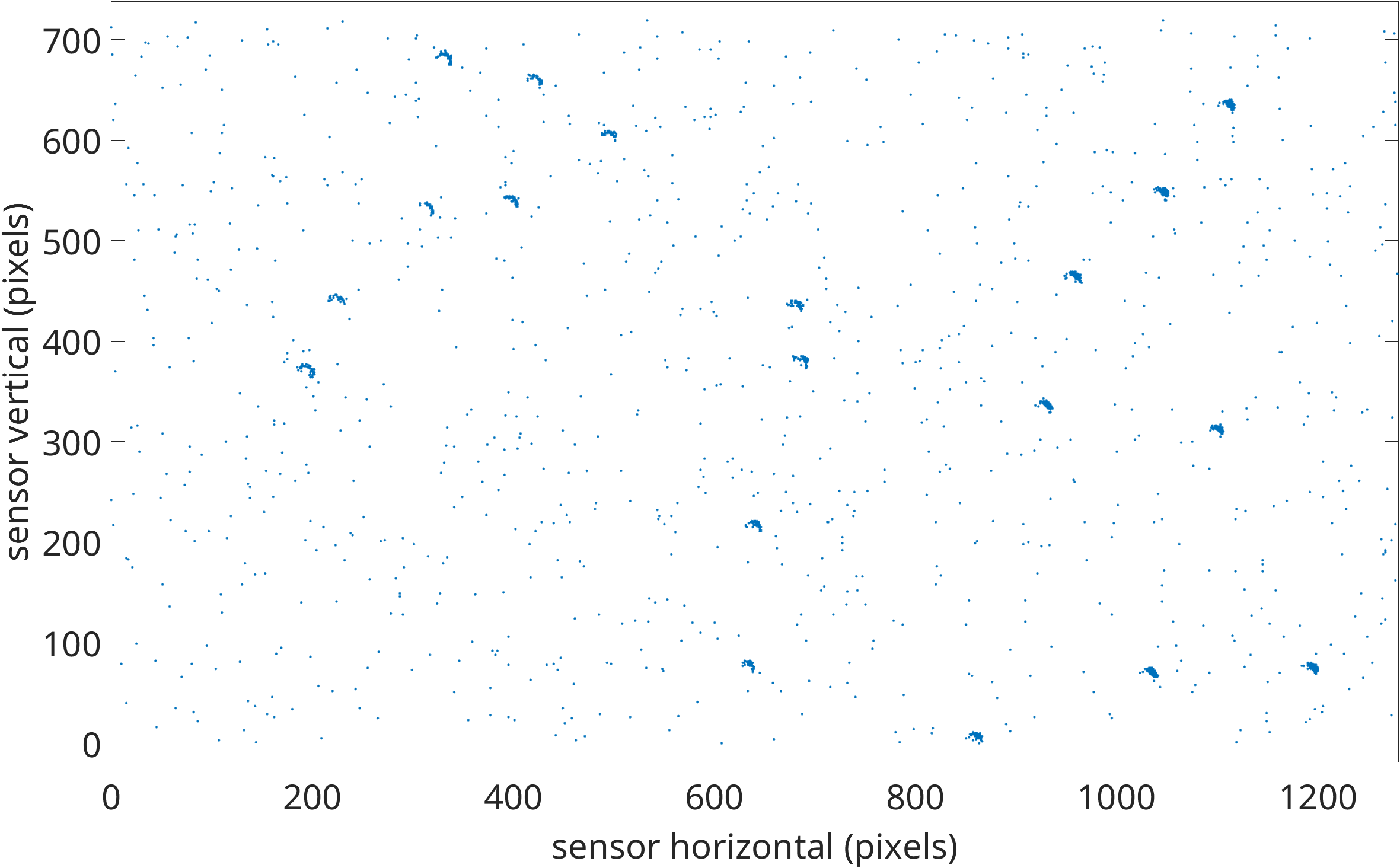}
    \includegraphics[width=0.49\textwidth]{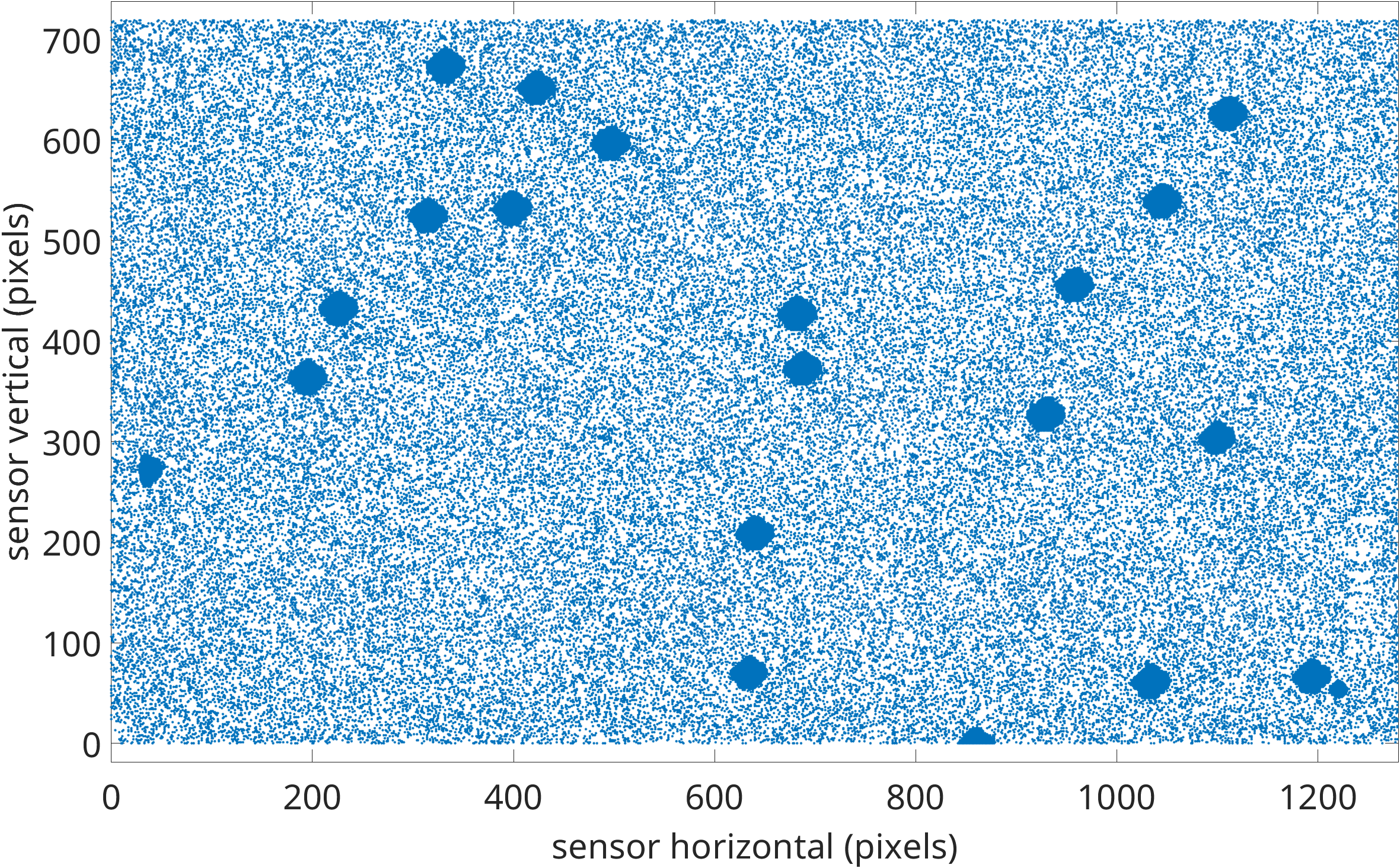}
    \label{fig:xyt-accumulation}}\\
    \caption{The sensing capabilities of an event sensor: (\textbf{a}): Space-time volume visualisation for a 1 second long event data stream when observing stars under high frequency perturbations. \textbf{(b)} The marked slices (red) spans 10 millisecond and are visualised as event frames in (c)(1-3). The event sensor provides high temporal resolution sensing for star positions even during this 10ms window. (c)(4) displays the perturbation observed over the complete second long exposure, as would be seen by a CMOS/CCD sensor. Long exposure leads to the light being smeared over a large number of pixels. Event sensing enables correction by offering star detection and tracking in the small timescale to prevent the smearing of photons in the longer timescale.}
    \label{fig:star-events}
\end{figure*}

Event sensors \cite{gallego2020event} offer a unique sensing modality that operates at high sampling rates, enabling characterisation of high-frequency perturbations. 
Compared to conventional cameras used on-board that operate in the 1-30Hz sampling rate (due to size, weight and power constraints) event cameras can sense changes in illumination asynchronously at 1MHz with lower power requirements while providing higher dynamic range. This combination provides a unique opportunity to use event sensors for on-board perturbation characterisation. Fig. \ref{fig:star-events} provides a visualisation of events during a 1 second window when observing at a star-field under perturbations. Streaks in the Fig. \ref{fig:xyt-volume} and \ref{fig:xt-slice} correspond to stars. Perturbations around a mean position can be observed even at such a small timescale (1 second). The event sensors generates high-frequency detection of the stars' positions on the sensor (Fig. \ref{fig:xyt-accumulation}), which can be used to rapidly detect any deviations from the desired pointing direction using the stars as a fixed reference. 

Using the high frequency sensing capabilities as the cornerstone, we design a payload that provides stable pointing over the exposure duration of the imaging sensor. Instead of relying on the ADCS to manoeuvre the whole space craft, 
we show that pointing errors can be reduced by stabilising the mission sensor \textit{independently} using an additional sensor -- an event sensor -- for standalone ``ultra-fine'' attitude estimation. We pair the low sampling rate (1-30Hz) mission sensor with a high sampling rate (kHz) event sensor to enable high frequency characterisation of the perturbations experienced by the satellite.
The proposed payload rapidly and accurately estimates the deviation from the required pointing attitude. Perception alone, however, can not provide the pointing stabilisation since software based post-processing of the main mission sensor will be rate limited by its sampling rate. Therefore, a physical actuation mechanism -- a micromotion stage -- is employed to keep the main mission sensor pointed towards the desired attitude. This is the second part of our contribution.
We refer to this mode of attitude estimation, where the residual correction from a desired attitude is computed via an additional sensors, as ``ultra-fine attitude estimation'' and the resulting attitude correct as ``ultra-fine pointing stabilisation''.  

% \begin{figure}[!t]
%     \centering
%     \includegraphics[width=0.7\textwidth]{figs/event-stream-stars.png}
%     \caption{Event stream generated by the event sensor when observing stars, visualised as a space-time volume. The thicker lines in the volume represent the observed stars. Notice the microsecond scale of the time axis.}
%     \label{fig:star-events}
% \end{figure}
%
%\textbf{Event sensor for SSA:}

In this paper, we present the design, algorithms and performance of a novel ultra-fine attitude estimation payload that takes advantage of the event-based neuromorphic sensor to provide ultra-fine attitude 
estimation for a co-located imaging sensor. The novelty of the proposed system comes from incorporating an event sensor for high frequency perception in the active stabilisation loop. This is combined with a piezoelectric motion stage, capable of providing repeatable motion in the $\eta$m range,  to execute high frequency corrections, together providing ultra-fine pointing stabilisation. 
The rest of the paper is organised as follows: Sec. \ref{sec:relatedwork} presents on overview of the prior art towards attitude estimation. Our system design is presented in Sec. \ref{sec:systemdesign} and the developed algorithms in Sec. \ref{sec:algorithms}. Evaluation of the system and performance results are presented in Sec \ref{sec:performance}.

\section{Related Work}\label{sec:relatedwork}
State-of-the-art commercial nanosatellite Attitude Determination and Control Systems (ADCS) claim a 1-$\sigma$ pointing precision in the order of tens of arcseconds (see Tab. \ref{tab:tracker-comp}). These ADCS solutions aim to stabilise the entire spacecraft and are prone to several sources of errors from actuation jitter, vibration of reaction wheels, control system update rate and latency, as well as external factors such as atmospheric drag. The module presented in work works independently of the ADCS to provide more precise attitude estimation and correction.

\begin{table}[h]
\centering
\begin{tabular}{c|c|c|c|c}
\hline
    Manufacturer & Model & Mass (g) & \multicolumn{1}{|p{3cm}|}{\centering Accuracy\\C/B-sight (arcseconds)} & Sampling rate (Hz)  \\\hline
    Adcole Space & MAI-SS & 170 &5.7 / 27 &4 \\\hline
    Blue Canyon Technologies    &   Standard NST    & 350   & 6 / 40       &    5     \\\hline
    CubeSpace                   &   CubeStar        &   55  & 77 / 220     &1          \\\hline
    Hyperion Technologies       &   ST200           &   42  & 30 / 200     &5          \\\cline{2-5}
                                &   ST400           &   280 &10 / 120    &5          \\\hline
    Jena-Optronik               & Astro APS         &   2000& 1 / 8        &16        \\\cline{2-5}
                                &   Astro CL        &   280 & 6 / 35       &10         \\\hline
    NewSpace Systems            &   NSGY-001        &  100  & 180 / 720    & 1        \\\hline
    Sinclair Interplanetary     &   ST-16RT2        & 185   & 5 / 55        &2        \\\hline
    Sodern                      & Auriga-CP         & 205   & 11 / 69       &10        \\\cline{2-5}
                                &   Hydra-CP        & 1400  &3.4 / 27       & 10     \\\hline
\end{tabular}
\caption{Comparison of various commercial star trackers in terms of accuracy and sampling rate. (Adapted from \cite{papotti2021star}). Astro-APS accuracy is provided at 1$\sigma$ while others are provided at 3$\sigma$}
\label{tab:tracker-comp}
\end{table}

Event sensors have recently been applied to space applications where they have been used for tasks such as domain adaption \cite{jawaid2023towards}, SSA \cite{cohen2019event} and object detection in space \cite{afshar2020event}.  The asynchronous nature of the event stream has also been taken into account to develop an asynchronous Kalman filter for star tracking using ground-based telescopes \cite{AKF}. Our proposal focuses on using the event sensor in an active stabilisation scenario, where the event sensors acts as an external source of information, alongside the main mission sensor for SSA. Such dual stabilisation has been successfully demonstrated in the ASTERIA mission \cite{pong2018orbit} using a combination of a CMOS sensor, used both for imaging as well as stabilisation, and a piezoelectric actuator for stabilisation. However, such a system can only compute the stabilisation corrections at the sampling rate of the CMOS sensor and will not be able to detect higher frequency perturbations.
Our proposal improves on their system by incorporating an additional event sensor and providing faster and more accurate stabilisation. The use of an additional sensor for attitude estimation task allows the CMOS sensor to focus solely on the task at hand of capturing the photons from the distant object of interest. From the perspective of the imaging sensor, this stabilisation is transparent.

\section{System Design}\label{sec:systemdesign}
\begin{figure}[!h]
    \centering
    \includegraphics[width=0.65\textwidth]{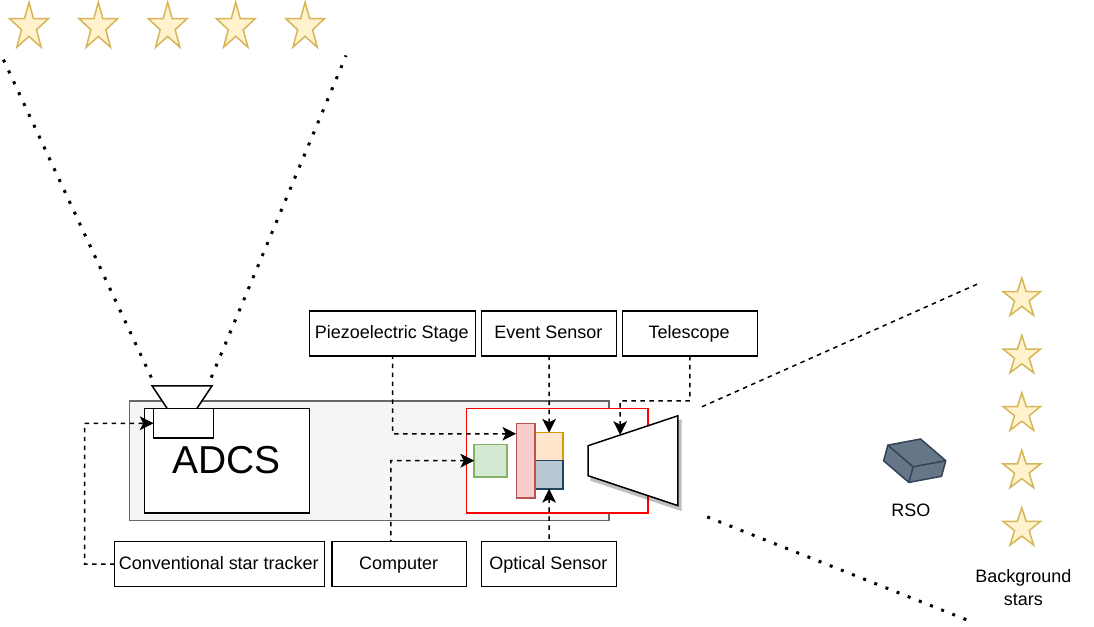}
    \caption{Operational context of the proposed ultra-fine attitude estimation and control payload. See text for details.}
    \label{fig:intro}
\end{figure}

%
% Active stabilisation uses background stars as fixed landmarks against which observed motion of the sensor is estimated for ultra-fine attitude estimation. The piezoelectric stage utilizes this estimation to minimise the remaining pointing error to allow ultra-fine motion compensation. The hardware utilised and algorithms that enable ultra-fine attitude estimation are presented in Sec. \ref{sec:hardware} and Sec. \ref{sec:algorithms} respectively.

\begin{figure}
    \centering
    \includegraphics[width=0.5\textwidth]{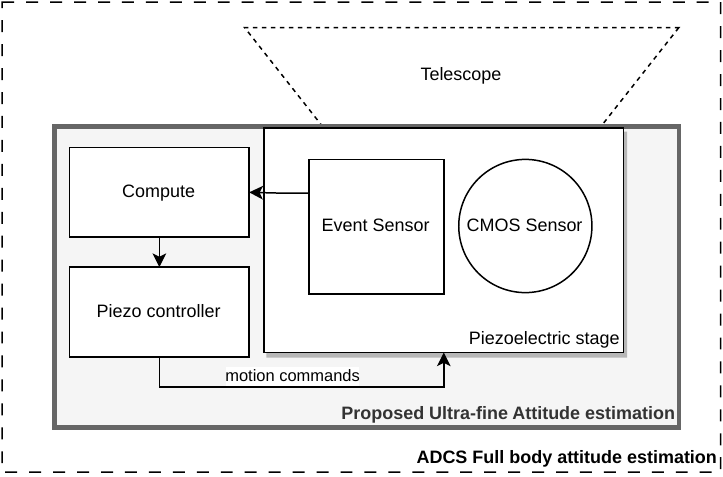}
    \caption{Proposed System Architecture: whole-body fine attitude control for the satellite is provided by the ADCS. The proposed ultra-fine attitude estimation is carried out via the event Sensor and ultra-fine control is carried out via the piezoelectric motion stage. The correction is computed by the on-board computer.}
    \label{fig:blockdiagram}
\end{figure}

%\subsection{Ultra-fine attitude estimation and control pipeline}

The operating context of the proposed payload is presented in Fig. \ref{fig:intro}. Background stars form a fixed reference against which instantaneous positions of the event sensor can be computed. The event sensor computes ultra-fine attitude at high frequency using this fixed reference. Discrepancy between the estimated and desired attitude is then minimised by driving a micromotion piezoelectric stage to provide high-accuracy pointing towards the desired Resident Space Object (RSO). The ADCS provides fine-attitude control while the ultra-fine adjustments are carried out using the proposed payload (marked in red) to reduce residual deviations from the required pointing direction.
Fig. \ref{fig:blockdiagram} depicts the high level architecture of the proposed payload designed to provide ultra-fine pointing stabilisation. The relationship between the onboard ADCS and the proposed module is also depicted in both. The ADCS provides fine attitude estimation and needs to manoeuvre the whole spacecraft to achieve pointing. The proposed payload sits within the body of the spacecraft and within the influence of the ADCS (as shown by the outer dotted region in Fig. \ref{fig:blockdiagram}). 
However, the task of the proposed module is to correct residual pointing errors in the system by employing its own perception (via the event-sensor) and actuation (using the piezoelectric motion stage) mechanisms. 
Light from the telescope is focused on the two sensors: the event-sensor and the optical CMOS sensor which is responsible for imaging the RSO of interest. 
 As both sensors are co-located on the piezoelectric motion stage, active stabilisation of the event sensor via the piezoelectric motion stage provides stabilisation for both sensor. This is how the proposed payload achieves stabilisation for the main mission (CMOS) sensor.  

  The onboard computer of the module receives data from the event sensor alongside the information about the desired attitude that needs to maintained during the image capture manoeuvre. Algorithms combine these pieces of information, compute the ultra-precise attitude of the sensor package, and generate motion commands for the piezoelectric stage, completing the action-perception loop and providing ultra-precise control. 
As part of the design, the optical sensor does not contribute to the stabilisation loop as it is the primary sensor for observations focuses solely on that task. 

% In short, EBPiezo provides an “active stabilization” platform which senses the discrepancy between the required and current pointing direction (after the fine control of ADCS has been applied) using the event-sensor and physically moves the sensors mounted on the piezoelectric platform to counteract the deviation, providing ultraprecise pointing for the sensor package. 

\subsection{Hardware}

In the section, we outline various hardware component of the payload, their main responsibility and how they are connected to other components.

\textbf{Event Sensor}
Event sensors  offer a novel sensing modality in the space domain. Unlike a conventional imaging sensor which captures an image at regular intervals, event-based sensing generates asynchronous ``events'' independently at pixels where the observed brightness changes. An event contains information about the pixel location, the direction of the observed intensity change and microsecond resolution timestamp of when the intensity change was observed.
%This allows microsecond resolution for the events, enables low latency attitude estimation. 
%Additionally, Event camera also offer also offer higher dynamic range compared to conventional cameras. 
% Event sensors for SSA
In the context of Space Situational Awareness (SSA), the scene is sparsely lit with vast swaths of black space and a few intermittent bright spots representing the stars. A conventional camera will spend energy in such a setting on repeatedly capturing the same region. 
Event-based sensors, in contrast, only generate events when intensity change is detected, either in the scene or induced by ego motion of the satellite. This leads to a smaller computation cost for downstream algorithms.
Event sensors are also well-suited to space based space observations as they offer a higher dynamic range compared to CMOS sensors, enabling operation in conditions where the CMOS sensors would either be over- or under-exposed, leading to a smaller sun exclusion angle. 

%As already described, the event sensor lies at the heart of the attitude estimation pipeline. 
Commercial off the shelf event sensors have made remarkable progress in terms of the number of pixels on the sensor as well as the reduction in the physical size of the sensor. Fundamental change in CMOS technology have allowed significant reduction in noise by stacking the light receiving and the processing circuitry on top of each other, leading to a smaller footprint for the sensor and a larger fill-in area. All have these advances have made event sensor more appealing to the task of SSA. 

% \yasir{Figure 11 Conventional vs. the new back-lit event sensor setup}

% \begin{table}[]
%     \centering
%     \begin{tabular}{c|c}
%     Property & Range \\\hline\hline
%     Resolution (pixels) & 1280 x 720\\\hline
%     Physical Chip size (mm$^2$) & 6.22 x 3.5 \\\hline
%     Pixel size ($\mu m^2$) & 4.86 x 4.86 \\\hline
%     Fill in factor (\%) & $>$ 77 \\\hline
%     Operational Voltage (V) & 1.1 and 2.5 \\\hline
%     Max. Bandwidth & 1066 Million event per second over USB3\\\hline
%     Angular FOV (100 mm optics, simulation) & 0.342 degrees x 0.241 degrees \\\hline
%     \end{tabular}
%     \caption{Caption}
%     \label{tab:sensor-specc}
% \end{table}

% In order to investigate the integrability of the event camera into our system, we show the internals of the Prophesee Gen 4 in Figure 12. The camera consists of stacked PCBs with the sensor at the top and the data-power connector at the bottom. This external housing consists of metal and contributes the most weight to the camera package. However, the internal PCB assembly is light enough to be mounted on the piezoelectric stage.

% \yasir{Figure 12 . Prophesee Gen4 internal: Left: Event sensor, middle back of the sensor package, right: The camera is composed of 4 PCBs connected to each other using the connectors marked in green. Each PCB is of the size 24mm x 19mm.}

\textbf{Piezoelectric micromotion stage}
To provide active stabilisation, the main mission sensor needs to be physically moved to make attitude corrections. This motion is generated by controlling a micromotion stage containing a piezoelectric motor as the actuator. Piezoelectric based motion mechanisms allow precise and repeatable motion execution. 

Two types of micromotion stages are found in practice: ``Stepper'' mechanism, where the micromotion of the piezoelectric component allows the mechanism to move a certain distance each time a voltage is applied. Such stages can hold their position without the need for additional power and do not need a homing mechanism, allowing the stage to carry out arbitrary motion sequences without the need to return to a known position (origin) between each motion step. However, stepper mechanisms tend to be comparatively slower, operating in the range of 10s of Hz with external motion input. 
The second class of piezoelectric motors are the ``compliant mechanisms'' based micromotion stages in which a compliant mechanism moves in response to the expansion and contraction of a piezoelectric element. Such motion stages are more responsive to input motion commands and can operate at much higher frequency (100s of Hz), however energy must be spent to keep the motor in a desired location. Additionally, such motors need homing -- returning to a known location, normally the origin -- between two motion commands. 

Based on lower power consumption and no need for homing, the proposed module uses a COTS stepper based piezoelectric motion stage that allows controllable and repeatable motion in the $\eta$m range. For completeness, the optics and onboard computer used for performance analysis of the developed prototype are described in Sec \ref{sec:performance}.

\section{Ultra-fine attitude stabilisation pipeline}\label{sec:algorithms}
\begin{figure}
    \centering
    \includegraphics[width=0.75\textwidth]{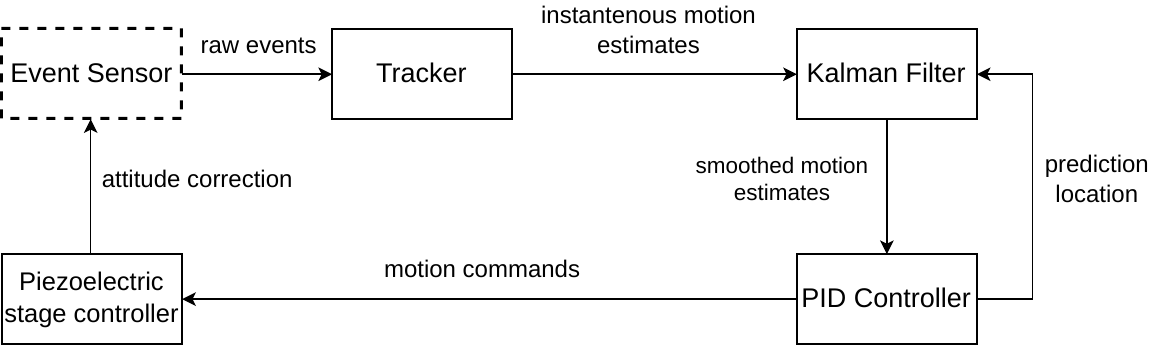}
    \caption{The algorithmic pipeline for the proposed system. The system converts input from the event camera into motion commands for the piezoelectric stage to actively stabilise the image position of the object of interest }
    \label{fig:algorithms}
\end{figure}

This section provides an overview of the algorithms that run on the on-board computer for ultra-fine attitude determination and control. Fig. \ref{fig:algorithms} shows the overall processing pipeline of the system. The main input to the system is ``Event sensor'' that represents the set of events $e_t$ observed by the event sensor. 
The tracker is responsible for generating instantaneous motion estimates (attitude) which are smoothed via Kalman Filter and passed to a PID controller to generate motion commands for the piezoelectric motion stage. In the following, we describe each of these modules in further detail.

\subsection{Input event stream}
The event stream consists of a list of asynchronous events $\mathbf{e}_t^i = [x_t^i, y_t^i, p_t^i, t^i]$, each containing the spatial location $[x_t^i,y_t^i]$ where an intensity change has been detected, a microsecond resolution time stamp $t^i$ and a polarity indicating $p_t^i$ whether the intensity at the location went up (positive) or down (negative) compared to the previous intensity at that location.
%For the testbed, these are events generated when the camera is pointed at the screen displaying the output of the star simulator. 
% The event stream is asynchronous, that is, unlike conventional cameras where each image is generated after a specific amount of time has elapsed, pixels in an event camera are triggered independently of each other. This effectively provides a high temporal resolution update of how the illumination in the scene is changing, leading to virtually zero motion blur in high-speed scenarios. Each event however contains very little information about the scene being observed. 
%
Instead of considering events individually, where each event provides very little information, we accumulate events for a fixed amount of time $\delta T$ into a ``batch'' and use these batches as an input to the tracking algorithm. Each batch consists of all the events within a time window:

\begin{equation}
\mathbf{B}_{T \to T + \delta T} = \{ \mathbf{e}_t^i~|~T  \le t^i < T + \delta T\}    
\end{equation}
 
Each batch is converted in a so called event-frame representation $F_{t}$ where the pixel in the frame is set to one for the location of the corresponding event in the batch.

\subsection{Tracker}
\label{sec:tracking}
\begin{figure}
    \centering
    \includegraphics[width=0.60\textwidth]{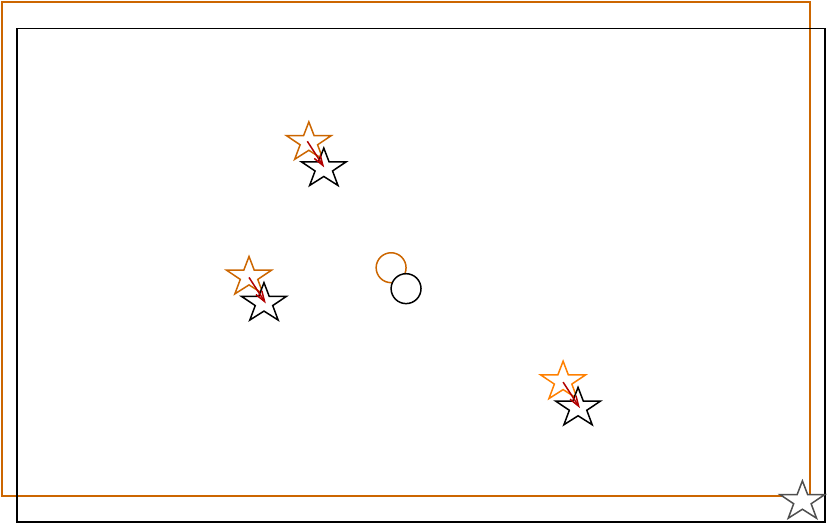}
    \caption{A schematic representation of the tracker sub-module for ultra-fine attitude estimation. Two consecutive frames $F_{t}$ (orange) and $F_{t+1}$ (black) are matched based on the detection and subsequent matching (red arrows) of the fixed stars in the background. This enables the estimation of motion between the two frames in the time between them. The accumulation time ($\delta T$) of the frame determines the frequency of the estimation. The bottom right corner depicts a star that has not been matched.}
    \label{fig:tracker}
\end{figure}
The tracking module is responsible for processing the input events frames and determining the amount of instantaneous motion between two consecutive event frames $F_t$ and $F_{t+1}$. This instantaneous motion estimate is used as input towards correcting the errors in the pointing direction. This module provides an estimate of motion at regular time intervals, determined by the accumulation time $\delta T$ of the batching step. 

For the tracking task, we exploit the properties of the scene being observed. The observed background stars are infinitely far away and within a very small enough time window, exhibit planar motion in the sensor space. Secondly, we can take advantage of the stars visible in the FOV of the sensor and use them as ``landmarks'' to compute the relative motion between two frames. Landmarks in this context means fixed observable entities that can be matched across time \cite{past-present-slam}. Therefore, for each event frame, we first isolate the location of bright stars -- circular clusters of pixels at least a few pixels wide (depicted as stars in Fig. \ref{fig:tracker}. We term this ``star detection''. The ability to successfully detect background stars is vital to the operation of the pipeline. Each of the $K$ stars detected in the $i$-the frame is represented by it centroid $\mathbf{s}^i_k, k = 1 \dots K$ consisting of its detected 2D location in the event frame.

As mentioned earlier, the event stream is not processed on a per-event basis, instead events are accumulated for a given duration to accumulate enough information for the subsequent star detection task. Once enough events have been accumulated, the algorithm detects a set of blobs -- clusters of bright pixels -- in the event stream. If no such blobs can be detected, enough information is currently not available to provide meaningful motion estimates and the system continue to wait until the next batch of events. When enough stars are detected for the first time, they are set as the ``origin'' of the system, against which motion estimates will be calculated for the future event frame. This forms a ``map'' of the sky containing stars that are locally visible around the current pointing direction. For each subsequent batch of events, stars are detected and aligned to the map already constructed, providing an instantaneous estimate of how much the system has moved since the starting position.

Mathematically, the task of the tracker is to align the set of stars in the current frame $\mathbf{s}^i_k$ against the $P$ star is map $\mathbf{s}^M_p, p = 1\dots P$. 
Given the underlying motion is a 2D translation, a least square estimate for the motion is obtained by first solving a corresponding problem (red arrows in  Fig.\ref{fig:tracker} to provide an initial set of matches between $\mathbf{s}^i_k$ and $\mathbf{s}^M_p$ using a nearest-neighbour matching technique using the current motion estimate $\textbf{t}_{i-1} \in \mathrm{R}^2$. Given the set of candidate correspondences $\mathbf{s}^i_k \to \mathbf{s}^M_{k'}$  from the currently observed stars to the map, the least squared estimated motion from the map for the current batch is computed as:
\begin{equation}
    \mathbf{t}_i = \frac{\sum_{k=1}^{K} ||\mathbf{s}^M_{k'} - \mathbf{s}^M_{k'}||_2}{K}  
\end{equation}
which is the average of the motion of the $K$ individual star motions. An important aspect of the problem, not shown in Fig. \ref{fig:tracker}, is the ``data association'' sub-module which find the most likely correspondence for the stars in the current frame against those in the map. We use a variant of the standard RANSAC algorithm \cite{RANSAC} reduce the effect of noise and outliers.

For reliable tracking, we require that at least 3 or more stars are present and matched at any moment in the system. When this condition fails, new stars that are already detected but not tracked are added to the map to allow the system to keep tracking robustly. This way, as the event sensor moves, new stars entering the field of view are added to the map and older stars which are no longer visible are removed from being tracked. This ensure that there are enough stars in the map at any given instance to provide robust tracking. Without active correction by the piezoelectric stage, this is an open loop estimate of the pointing direction of the satellite. 

Of special interest to the present task is the ability of the event sensor to asynchronously provide events at the rate of 1MHz. This allows rapid change detection in the scene caused by high frequency perturbations experienced by the satellite. We report results for update rate of 100Hz in Sec. \ref{sec:performance}.

\subsection{Kalman filtering}
As the exposure time for the main sensor increases, the ADCS will execute a capture manoeuvre to keep the RSO in sight. In this case, there is a low-frequency signal that needs to be tracked, buried inside the high-frequency perturbations. The instantaneous motion estimates are noisy and unaware of the trajectory being executed by the ADCS. We resolve this by employing the Kalman filter \cite{KF} which introduces prior knowledge about the motion of interest. This allows smoothing the motion estimates over time to recover the low frequency underlying signal. Additionally, the instantaneous motion estimate only provides an estimate of displacement (change in position) but for motion planning we need an estimate of the velocity of the system, which is not directly observable. The Kalman filter can take in the observed variables (displacements) and computes hidden state variable (velocity) from it over time.
% We employ a Kalman Filter to both smooth out the motion estimates and provide an estimate of the velocity of the system. 
%
The state of the Kalman filter contains the position and velocity of the event sensor:
\begin{equation}
    \mathbf{x}_k = [p_x~p_y~v_x~v_y]^{T}
\end{equation}
which evolves over time as 
\begin{equation}
    \mathbf{x}_{k+1} = \mathbf{Fx}_{k}  + \mathbf{Bu}_{k+1} + \mathbf{w}_k
\end{equation}
The state transition matrix $F$ is assumed constant over time and represent the constant velocity motion in $\mathbb{R}^2$
\begin{equation}
    \mathbf{F} = \left[ \begin{array}{cccc}
         1&0&\delta T & 0 \\
         0& 1 & 0 & \delta T\\
         0&0  &1  &0 \\
         0&0  &0  &1 \\
    \end{array}
    \right]
\end{equation}
The control input $\mathbf{u}_{k+1}$ is computed as a motion command via the PID controller (see next section) and fed back to the KF for integration. $\mathbf{B}$ is therefore the $\mathrm{I}^{2 \times 2}$ matrix. Finally, $\mathbf{w}_k$ is the additive process noise.
The assumed constant velocity model is updated on each computed motion estimate. The filter balances what we observe (instantaneous motion) and what we believe to be true about the operating conditions (constant velocity). The computed estimate of the velocity is used to predict the position of the event sensor at the next time instance by the motion controller.

\begin{figure}
    \centering
    \includegraphics[width=0.49\textwidth]{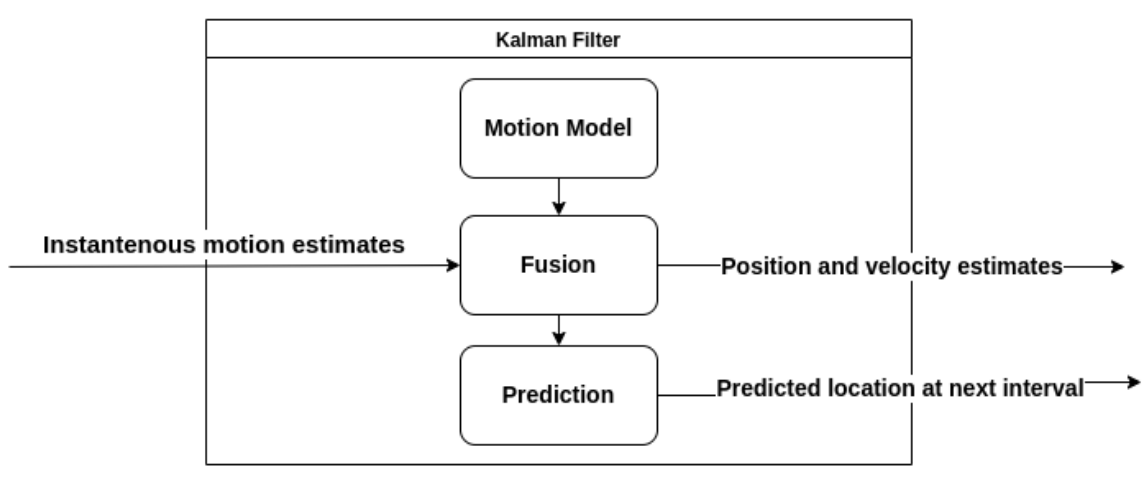}
    \includegraphics[width=0.50\textwidth]{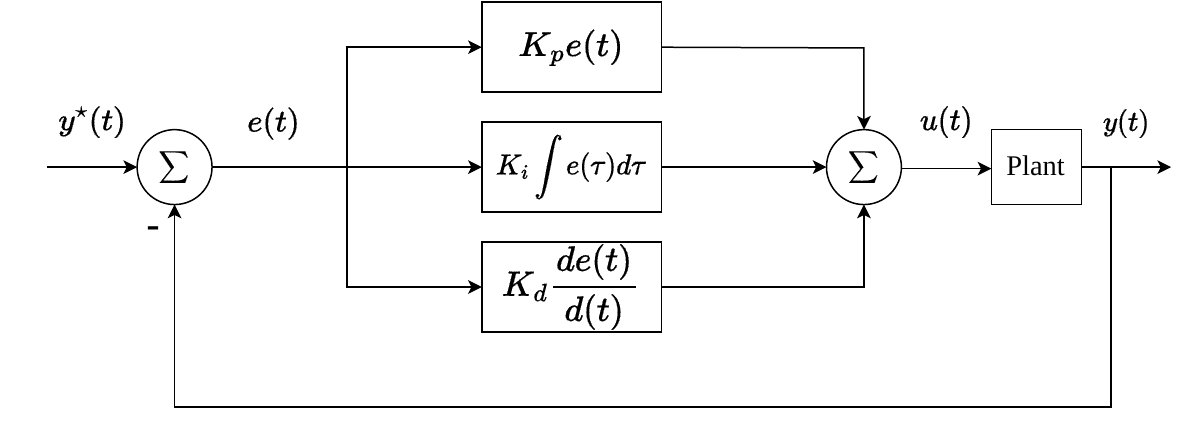}
    \caption{\textbf{Left: }Kalman Filtering to Smooth and Predict the state of the system \textbf{right :} A PID controller driving the system to desired state $\mathbf{y}^{\star}(t)$ by minimising its difference to the current state $\mathbf{y}(t)$ }
    \label{fig:KF-PID}
\end{figure}

\subsection{Stabilisation using Piezoelectric actuation}
PID controllers \cite{PIDController} are a standard device from control literature to drive a system to a desired state and works by minimising the error between the current and the target location using Proportional, Integral and Derivative (PID) error terms (Fig. \ref{fig:KF-PID}). 
Given a desired state, $\mathbf{y}^{\star}(t)$, the controller computes the correction that needs to be applied based on the estimated error $\mathbf{e}(t)$:
\begin{equation}
    u(t) = K_p e(t) + K_i \int e(\tau) d\tau + K_d \frac{de(t)}{dt}
\end{equation}
where $\mathbf{e}(t) = \mathbf{y}^{\star}(t) -\mathbf{y}(t)$ is the discrepancy between the desired and current state.
The contribution of each of the terms is weighted using the constants $K_{\star}$. The proportional term $K_p$ reacts linear to the error term while the integral  $K_i$ term operates on the residual error in the system over time. The derivative $K_d$ term follows the current gradient of the error for future corrections. The PID controller output $\mathbf{u}(t)$ instructs the piezoelectric stage to move in the desired pointing direction by driving the error to zero over time. The current position $\mathbf{y}(t)$ is obtained from the Kalman Filter.

% \subsection{High Frequency Tracking}
% The tracking system described until now operates at 10Hz by accumulating information over 100ms from the event camera. This is done to comply with the operational frequency of the piezoelectric motion stage. However, the open-loop tracking can operate at a much high frequency. To illustrate this, we have implemented a high frequency intermediate tracker, which operate within the 100ms interval while data is being collected for the main tracking algorithm.
% This tracker integrates instantaneous motion hypothesis by looking at events within a 10ms interval (100Hz operational rate). During this interval, information about observed motion is computed and is used to inform the main tracking algorithm. Such an algorithm is useful in high magnitude jitter cases when the observed motion can be significant in smaller time intervals. We demonstrate the performance of high-speed tracking in the experiment section.

\section{Performance Analysis}\label{sec:performance}
\begin{figure}
    \centering
    \includegraphics[width=0.59\textwidth]{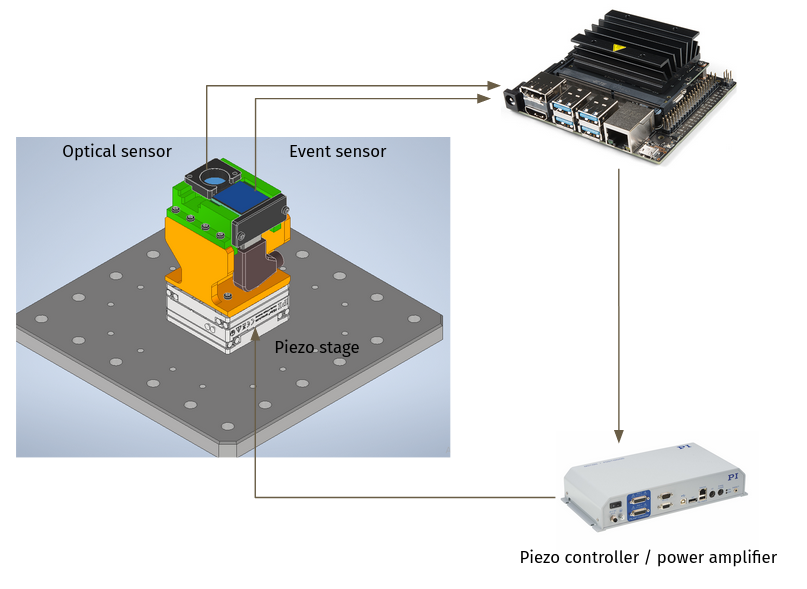}
    \includegraphics[width=0.35\textwidth]{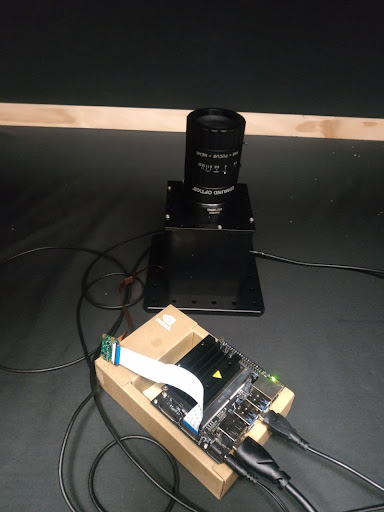}
    \caption{(Left) CAD model of the developed sensor package, compute and piezoelectric stage controller (Right) The developed prototype shown in the dark room}
    \label{fig:testbed}
\end{figure}

In this section, we first describe the developed prototype and the conditions in which the performance analysis is performed. 
We report performance for the attitude estimation in an open-loop configuration at different operating frequencies. We also show the performance for the closed loop stabilisation task. We conclude the section with a discussion about execution time and its dependence on the hardware. 

\subsection{Developed Prototype}
A prototype of the proposed module has been developed using commercial-off-the shelf (COTS) components, which replicates the full system as closely as possible while still being testable within the laboratory conditions. 

\textbf{Optical setup}
The in-space module will sit within the light cone of a telescope, however such an optical setup is not testable within laboratory conditions. 
Therefore, the optics have been simulated using a conjunction of 100mm lens and a display screen to simulate stars at the correct scale. The optical focal plane sits at a more reasonable distance of 2m from the sensor plane. With the current setup, the event sensor has a (simulated) angular field of view of $1.5^\circ \times 0.8^\circ$. %resulting in AFOV of $0.4212 \times 0.4212$ arcseconds for each pixel. 

\textbf{Sensor package}
The prototype contains the event sensor and an optical sensor inside a custom made housing. The micromotion stage is  at the bottom of the structure, to which the sensors are attached via a riser (Fig. \ref{fig:testbed}). This riser allows input-output connectivity via the original sensor evaluation kit connectors. A stabiliser plate connects the micromotion stage to the base of the housing. 
The custom-built enclosure allows for the optics to be placed in the correct position relative to the focal plane of the optics. Cut-outs within this enclosure allow routing cables from the sensors to the computer and to the piezoeletric motion stage. 

\textbf{Compute}
An onboard computer that processes the sensory information into an estimation of motion for the micro-motion stage to execute. The sensors are connected to the onboard computer, which for the case of the prototype consists of a Jetson Nano\footnote{\texttt{https://developer.nvidia.com/embedded/jetson-nano-developer-kit}} development board.

\begin{figure}
    \centering
    \subfloat[][Schematics of the dark room.]{    
\fbox{
\includegraphics[width=0.22\textwidth,valign=b]{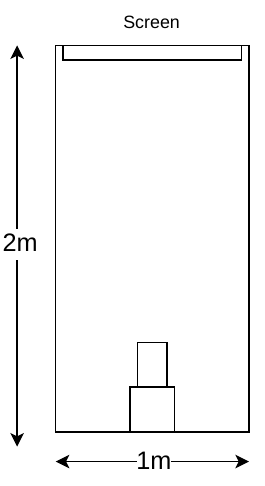}}
    \label{fig:darkroom}
    }\quad
\subfloat[][Star Simulator's output displaying a sparse star field (inverted colour. Best seen digitally.)]{    
    \fbox{\includegraphics[width=0.7\textwidth,valign=b]{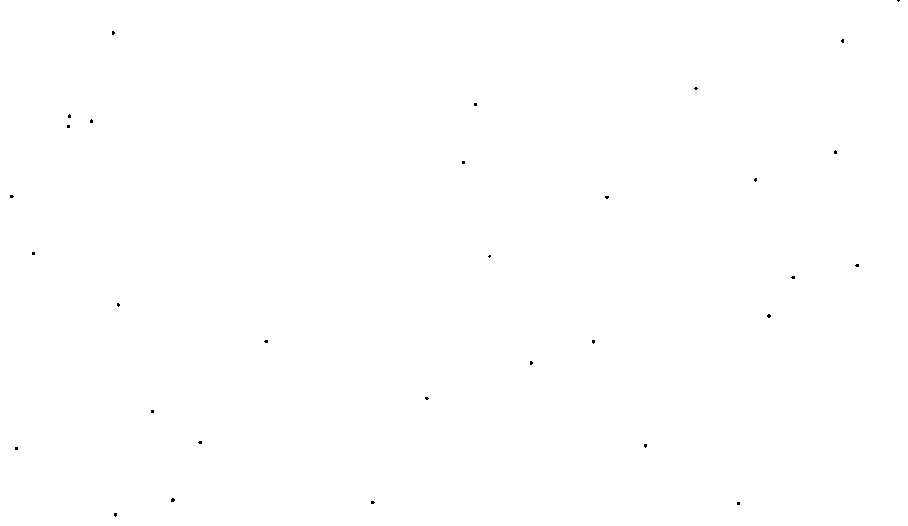}}
    \label{fig:simulator-output}
    }
    \caption{Test bed setup showing the dark room and simulator output.}
\end{figure}

\subsection{Test bed design}
To benchmark the module in a laboratory setting, it is essential to have a test environment that replicates the on-board observations with fidelity. We achieve this by generating simulated stars and housing the prototype in a dark room, both are which are described in the following.

\textbf{Star simulator}
To simulate the appearance of stars and objects of interest, a custom GPU based star simulator has been developed which reads positions of bright stars from the Tycho2 catalogue \cite{tycho2} and displays them on a high refresh rate screen. The simulator is highly configurable and can simulated various fields of view, the initial pointing directions and frequencies of image generation. Only stars brighter than a certain visual magnitude are included. A sample image generate by the simulator (inverted so that stars appear dark) is shown in Fig. \ref{fig:simulator-output}.
In addition to displaying an input star field to the system,
the simulator is used to collect fine grained ground truth motion information which is used to benchmark the performance of the system. 

\textbf{Dark room}
To prevent interference from stray light, the system is contained inside a darkroom as shown in Fig.\ref{fig:darkroom}. It consists of a custom-made 2-metre-tall wooden structure with a base of 1m x 1m and houses the system prototype as well as a display screen which has a resolution of 1920 x 1080 and is capable of a refresh rate of 120Hz. It is used to display the images generated by the star simulator.

% \begin{figure}
%     \centering
% \includegraphics{figs/draw_room.pdf}
%     \caption{Schematic of the dark room where the prototype is benchmarked.}
%     \label{fig:darkroom}
% \end{figure}

%\yasir{Figure 8 Dark room layout}

\subsection{Ground Truth acquisition}
An important aspect of system evaluation is the ability to collect reliable and representative ground truth data to certify system performance. To achieve this, the star simulator is used to generate various motion patterns of the star field as observed by the sensor package instead of moving the developed prototype at high frequencies which is technically challenging. 
We simulate two kinds of trajectories: a) benchmarking trajectories test the system performance under harsh condition while b) realistic trajectories closely depict the expected on-board observations. 
The benchmarking trajectories include a \texttt{circle} and \texttt{square} trajectory. 
Realistic trajectories are based \texttt{linear} motion but with added noise at various operational frequencies. 
Table \ref{tab:trajectories} provides further details about these trajectories.
\begin{table}[!h]
    \centering
    \begin{tabular}{c|l}
    Trajectory & Properties \\ \hline\hline
    Square  & Side length: 0.1 degrees (360 arcsec) \\ \hline
            & Execution time: 20 seconds \\\hline
    Circle  & Radius: 0.05 degrees (180 arcsec) \\\hline
            & Execution time: 45 seconds \\\hline
    Linear & Velocity: 0.005 degrees / seconds (18 arcsec / second) \\ \hline
    \end{tabular}
    \caption{Benchmarking trajectories used in the simulator}
    \label{tab:trajectories}
\end{table}

The ground truth from the simulator consists of timestamped attitude in as Right Ascension (Ra) and Declination (Dec) for each generated frame, along with a system timestamp. Timestamps are used to align the estimates against the ground truth. 
%
%In view of better understanding how the process affects light capture at the pixel level, ground data positions are converted to and aligned with the event sensor pixels and results are reported in pixel space of the event sensor. 

System performance under perturbation is evaluated by simulating additive zero-mean Gaussian noise with increasing variance over an exponential scale ($ \sigma_N = 10^{-9}$ to $10^{-6}$ degrees).   
%To explore the performance of the developed algorithms and hardware under high frequency perturbations, we simulate Gaussian noise of varying magnitudes at an exponential scale. 
This high-frequency (30 Hz to 100Hz) additive noise corrupts each incremental motion estimate. 
% /The noise is drawn from Gaussian distributions with standard deviation from $ \sigma_N = 10^{-9}$ to $10^{-6}$ degrees translating to pixel offset with magnitudes of 0.36 to 20 pixels in the event sensor space respectively. 
% This gaussian noise is injected at each step of the simulator (each frame) by adding the perturbation to the ground truth trajectory. 
% The expected noise in the real system is in the 1e-9 to 1e-8 range. Higher noise levels serve as stress test for the system. 

\subsection{Tracking performance}
Pointing accuracy relies heavily on accurate ultra-fine attitude estimation. We first evaluate the \textit{open loop} operating setting in which the estimated attitude is not applied for correction. This enables evaluation of the ultra-fine attitude estimation sub-task. When the estimate drives the piezoelectric stage, we term it \textit{closed loop} stabilisation.

% and enables stabilization by providing an up-to-date estimate of the scene being observed against the position of event sensor. 
%This is the first step in the stabilisation pipeline since the estimated motion serves as the basis for the corrective motion executed by the motion stage. 

\subsubsection{Open Loop High Frequency Tracking} \label{sec:hf}
% \begin{figure}
%     \centering
%     \includegraphics[width=0.45\textwidth]{figs/t_noise_1e_6_full.png} \\
%     \caption{Caption}
%     \label{fig:high-freq-results}
% \end{figure}

\begin{figure}
     \centering
     \subfloat[][Demonstrative results for high-frequency tracking at $\sigma = 10^{-6}$ degrees. The first column demonstrates the ground truth trajectory in red and the estimated trajectory in blue. The subsequent columns show the regions marked in red, green and blue respectively in the first column.]{
     \includegraphics[width=0.28\textwidth]{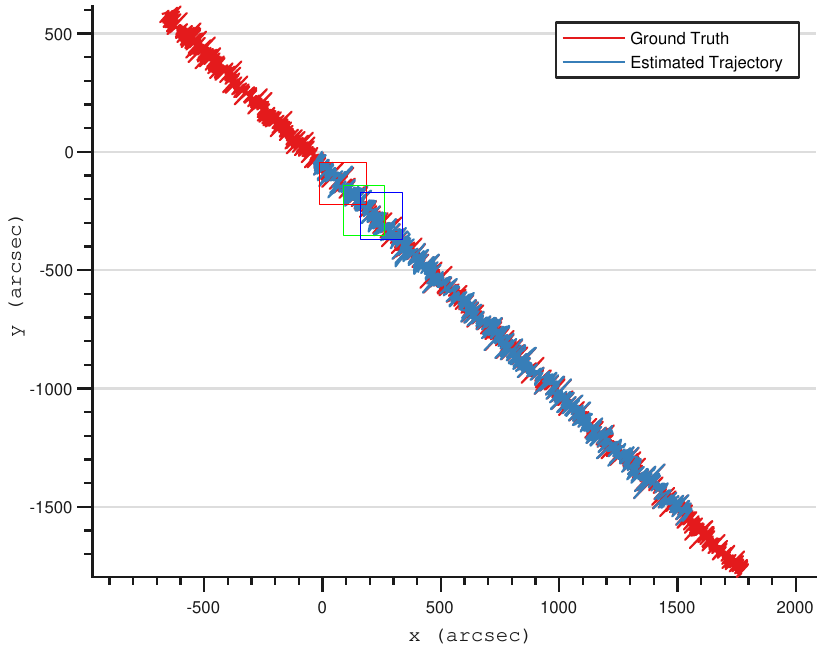}
     \includegraphics[width=0.23\textwidth]{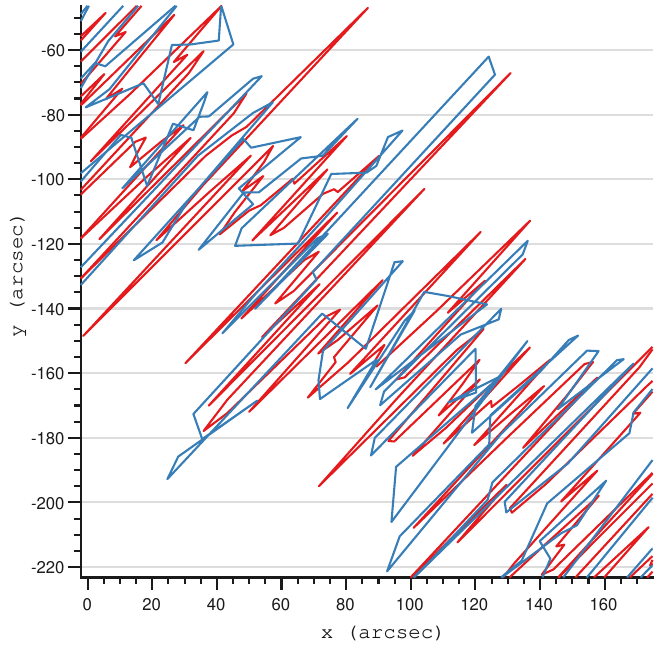}
     \includegraphics[width=0.23\textwidth]{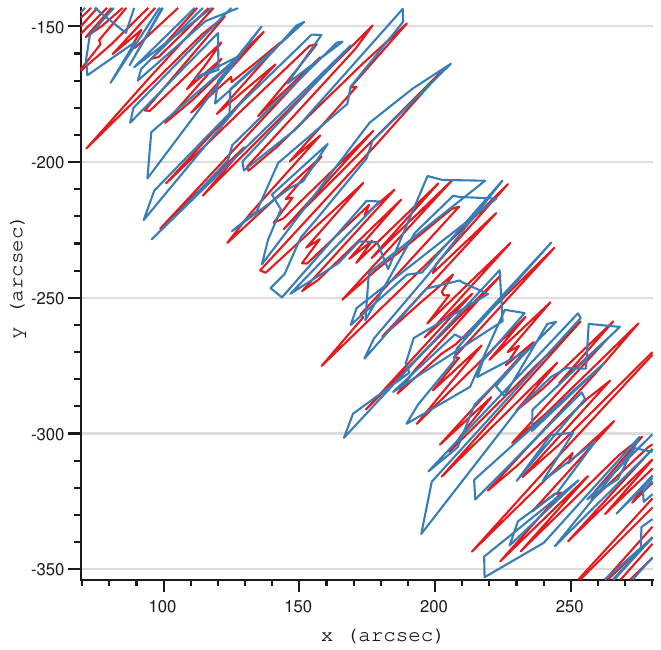}
     \includegraphics[width=0.23\textwidth]{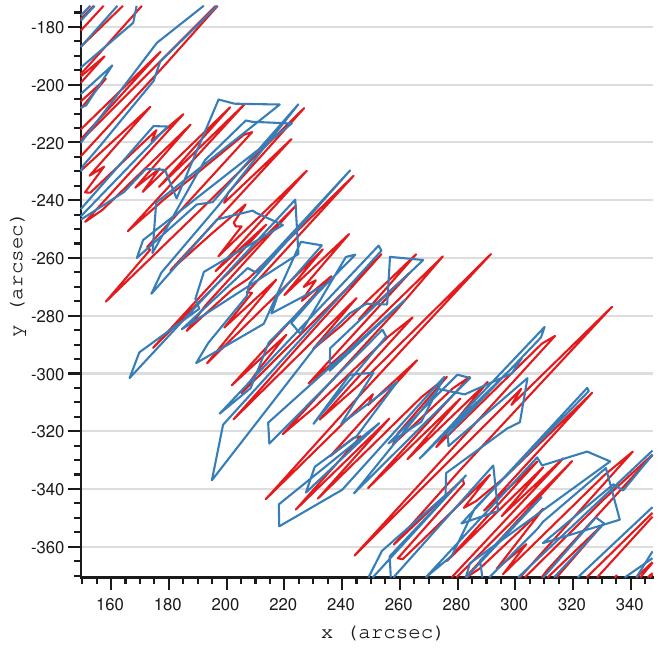}
     \label{fig:highf_insets}
     } \\
     \subfloat[][Tracking performance along the x-direction. \textbf{Right}: A zoomed in view starting at t = 3 seconds along the x-direction.]{
     \includegraphics[width=0.45\textwidth]{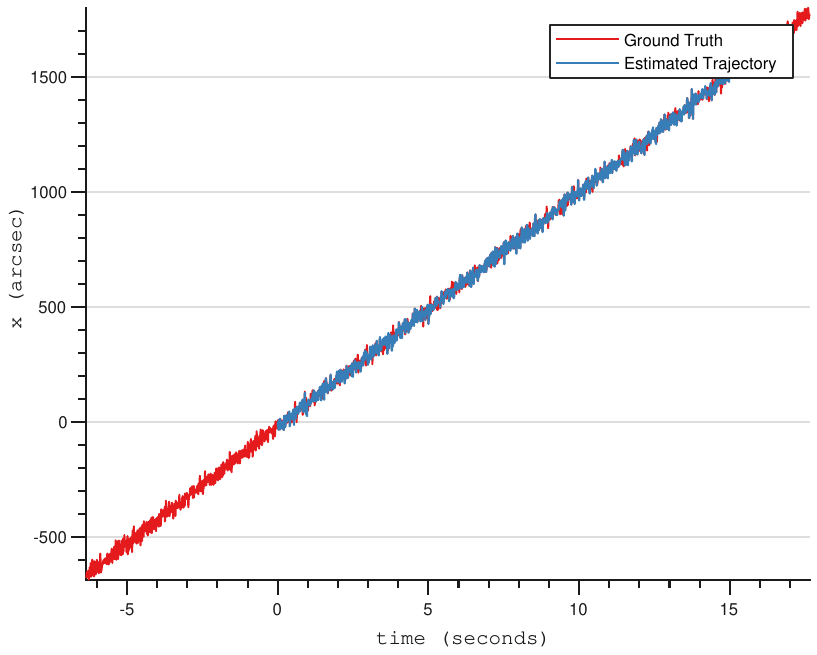}
     \includegraphics[width=0.45\textwidth]{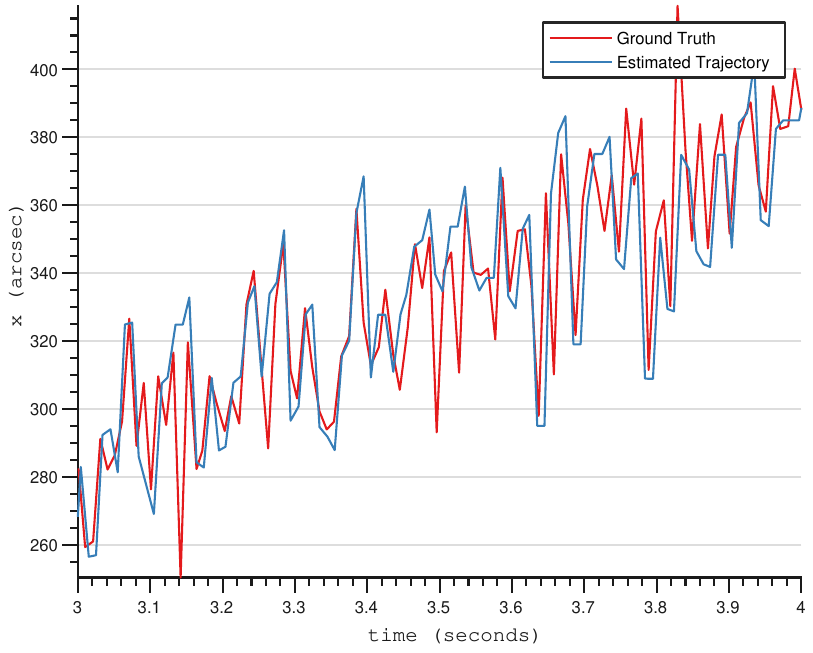} }\\
     \subfloat[][Tracking performance along the y-direction. \textbf{Right}: A zoomed in view starting at t = 3 seconds along the y-direction.]{
     \includegraphics[width=0.45\textwidth]{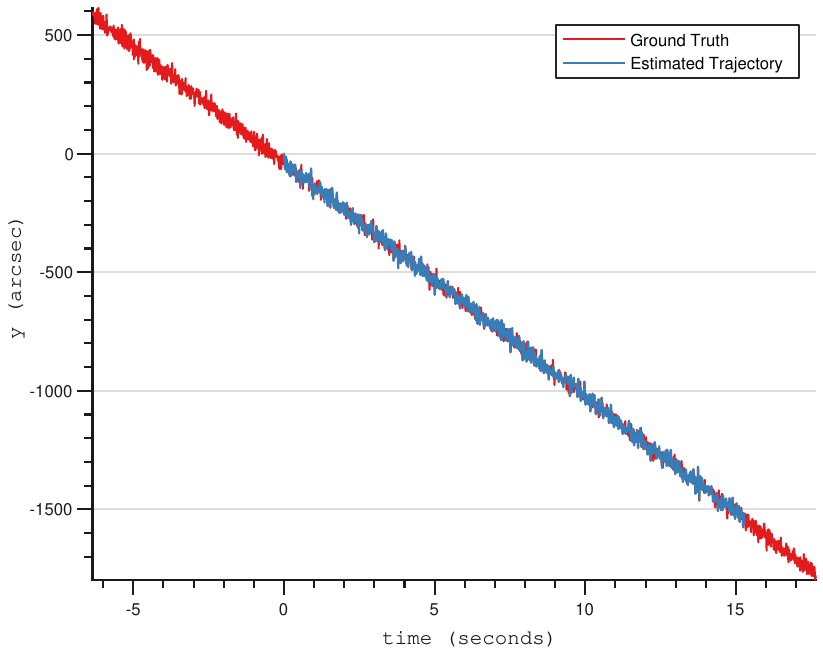}
     \includegraphics[width=0.45\textwidth]{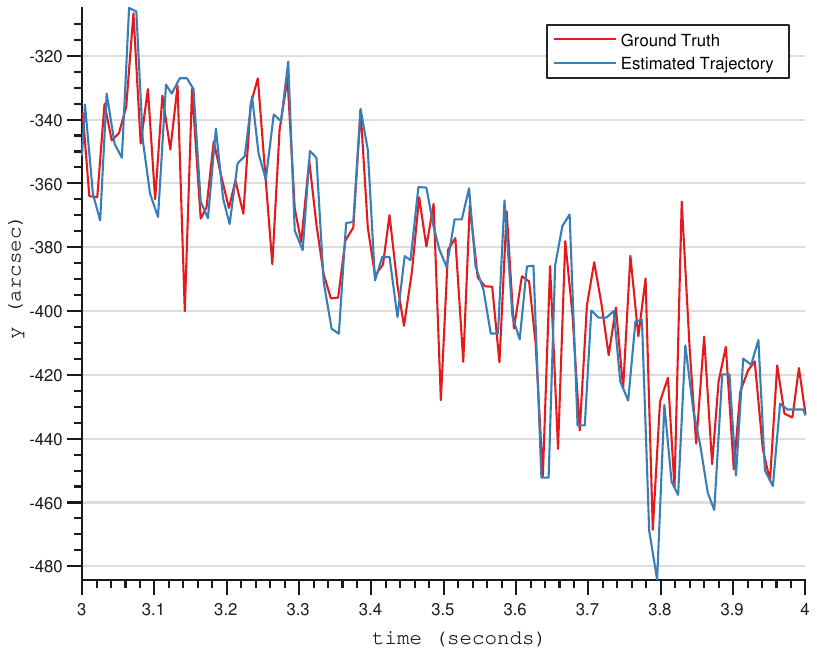} 
     } \\
     \caption{Results for high-frequency tracking}
     \label{fig:high-freq-results}
\end{figure}

% \begin{figure}
%      \centering
%      \subfloat[][Demonstrative results for high-frequency tracking at $\sigma = 10^{-6}$ degrees. The first column demonstrates the ground truth trajectory in red and the estimated trajectory in blue. The subsequent columns show the regions marked in red, green and blue respectively in the first column.]{
%      \includegraphics[width=0.28\textwidth]{figs/results/hf_main_6.png}
%      \includegraphics[width=0.23\textwidth]{figs/results/hf_main_6_1.png}
%      \includegraphics[width=0.23\textwidth]{figs/results/hf_main_6_2.png}
%      \includegraphics[width=0.23\textwidth]{figs/results/hf_main_6_3.png}
%      \label{fig:highf_insets}
%      } \\
%      \subfloat[][Tracking performance between t = 3 and t = 4 seconds. Top and Bottom rows show result of tracking along the x- and y-directions respectively.]{
%      \includegraphics[width=0.99\textwidth]{figs/t_noise_1e_6_time_full.png}
%      } \\
%      \caption{Results for high-frequency tracking}
%      \label{fig:high-freq-results}
% \end{figure}

% \begin{figure}
%     \centering
%     \includegraphics[width=0.6\textwidth]{figs/t_noise_1e_6_time_full.png}
%     \caption{Tracking performance between t = 3 and t = 4 seconds. Top and Bottom rows show result of x- and y-direction respectively against time.}
%     \label{fig:high-freq-results-time}
% \end{figure}
We first consider the case of high-frequency relative attitude estimation where the satellite experiences high-frequency (100Hz) high-magnitude jitter. This type of jitter can be caused by factors including space weather and atmospheric drag.
%
%The experiment is designed to demonstrate the ability of the proposed method to work at an operational frequency of the noise (100 Hz in this case) along with the ability to track instantaneous motion at a much higher magnitude than observed in space. 
%
To enable the high-speed tracking mode, we utilise only the tracking module (Sec \ref{sec:tracking}) without the Kalman filter. This frees up the motion prediction from any assumption on the motion prior enabling high frequency tracking of the signal along with the injected noise.  We benchmark how quickly and accurately the noisy signal can be tracked without the corresponding smoothing effect introduced by the Kalman Filter. This allows us to analyse the direct effect of tracking module.
The simulator generates Gaussian noise with standard deviation of  $\sigma_N = 10^{-6}$ degrees leading to over 84 arcseconds in the event sensor view \textit{per} incremental motion step, simulating high magnitude jitter. 
It should be noted that at such high noise levels, the magnitude of the jitter is far higher than the expected underlying motion. Therefore, it is sufficient to demonstrate the tracking performance of the system on the \texttt{linear} case as all trajectories are approximately linear in the small time window.

Fig. \ref{fig:highf_insets} demonstrates the estimated trajectory (blue) and against the ground truth (red) for the high frequency tracking experiments. The noise levels experienced by algorithm can be seen in insets (columns 2-4). It can be seen that as the algorithm tracks the jitter, it also tracks the underlying signal. To better view the tracking performance, we show separate plots for tracking along the x- and y- direction against the tracking time in Fig. \ref{fig:high-freq-results}b with zoomed version for a short time in the right column. Thanks to the event sensor's high sampling rate, we can accurately track the jitter both in the x- and y- directions while tracking the overall signal with perturbations greater than 10 arcseconds frequently seen in the plots at each timestamp.
These plots demonstrate the efficacy the event sensor in tackling high frequency, high magnitude jitter. A high frequency actuation mechanism can be used to compensate for the computed motion at provides stabilisation. 

\subsubsection{Open Loop Tracking Accuracy}
We focus on the comparatively low frequency (10Hz) case  used for stable pointing. Stabilisation via the micromotion stage needs depends on accurately estimating of the sensor's position. This raw position estimate is smoothed  via the Kalman filter to provide a consistent and reliable estimate of the sensor's position for the stabilisation task. Experiments in this section track stars in the event stream at $10$Hz by accumulating event data for $100$ms to generate event frames. 

\begin{figure}
    \centering
    \includegraphics[width=0.5\textwidth]{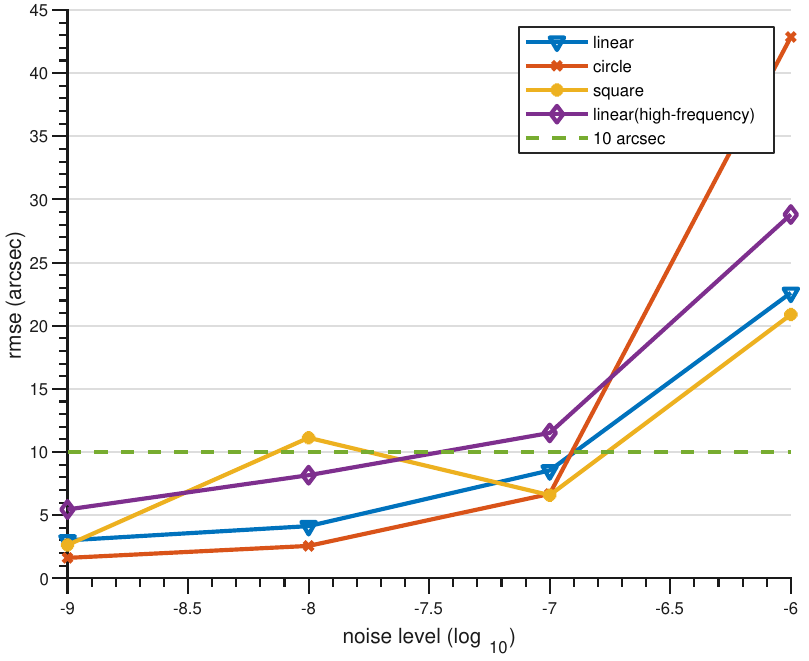}
    \caption{Result for simulated trajectories and performance results for various noise levels.}
    \label{fig:rmse}
\end{figure}

\begin{figure*}
     \centering
     \subfloat[][\texttt{linear} trajectory]{
     \includegraphics[width=0.28\textwidth]{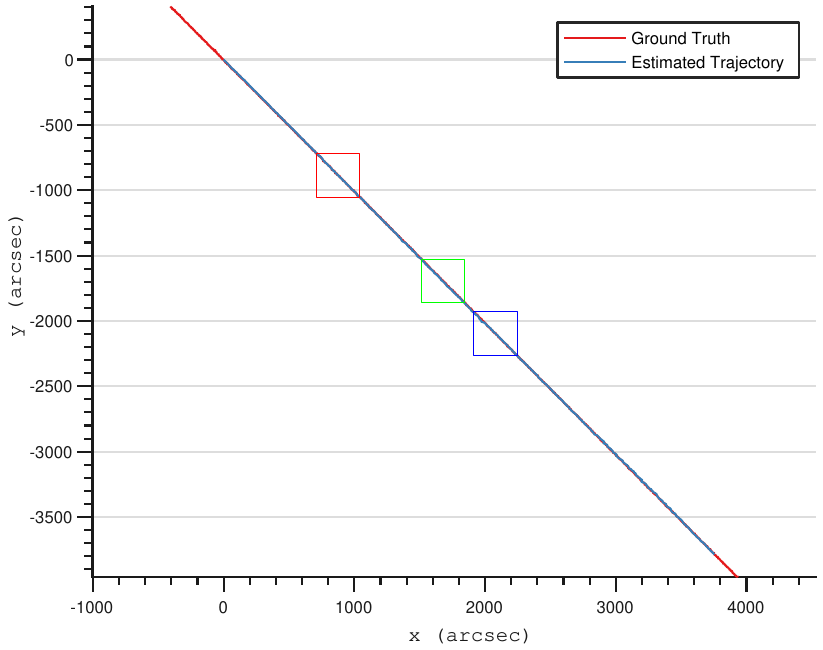}
     \includegraphics[width=0.23\textwidth]{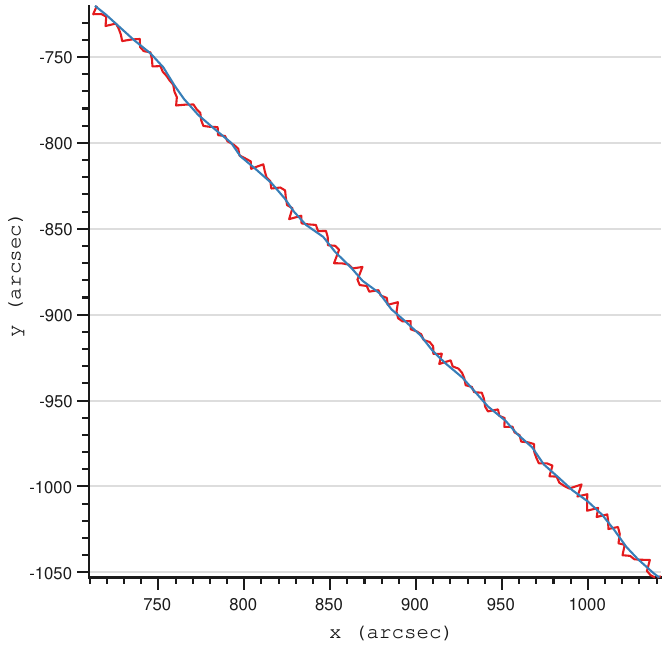}
     \includegraphics[width=0.23\textwidth]{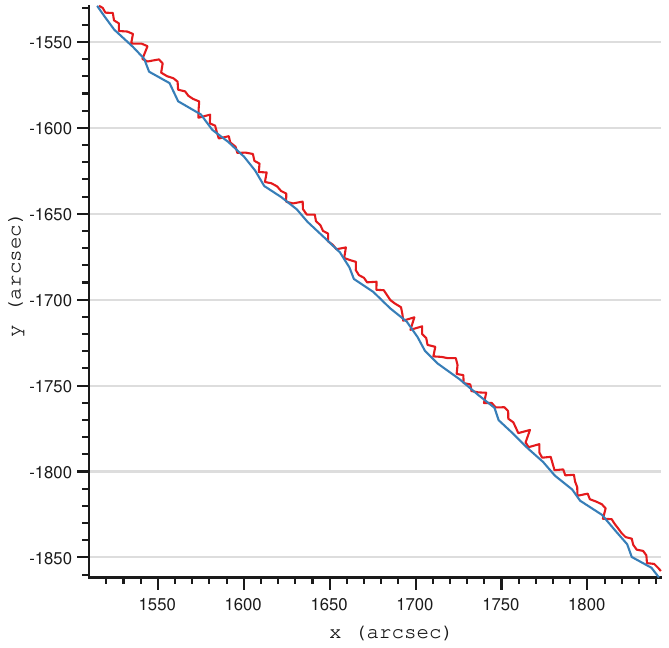}
     \includegraphics[width=0.23\textwidth]{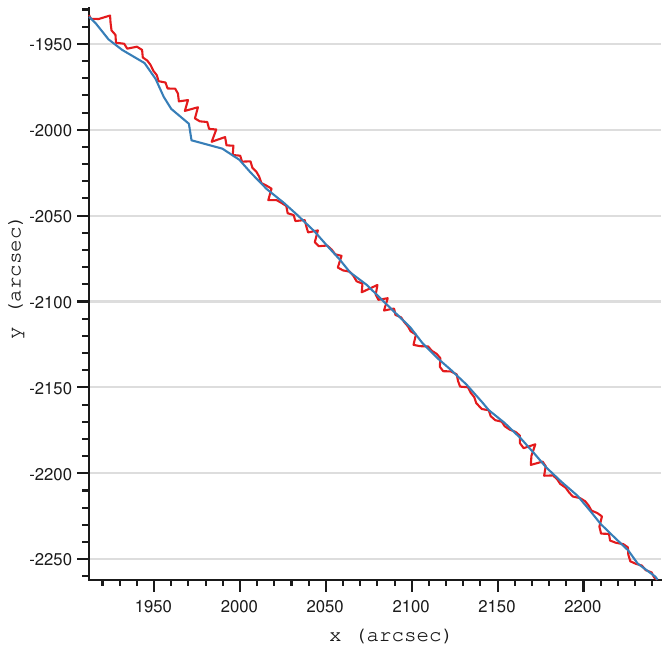}
     } \\
     \subfloat[][\texttt{square} trajectory]{
     \includegraphics[width=0.28\textwidth]{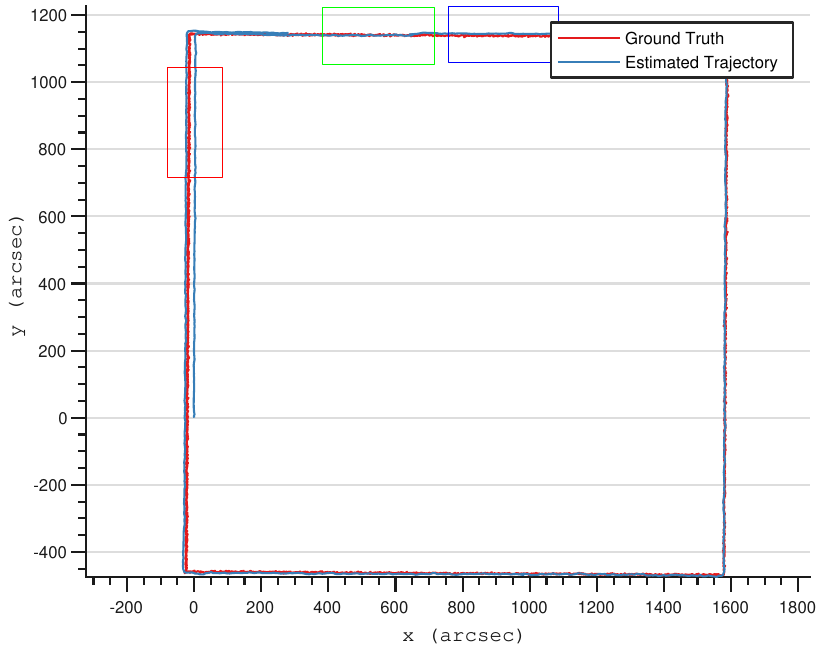}
     \includegraphics[width=0.23\textwidth]{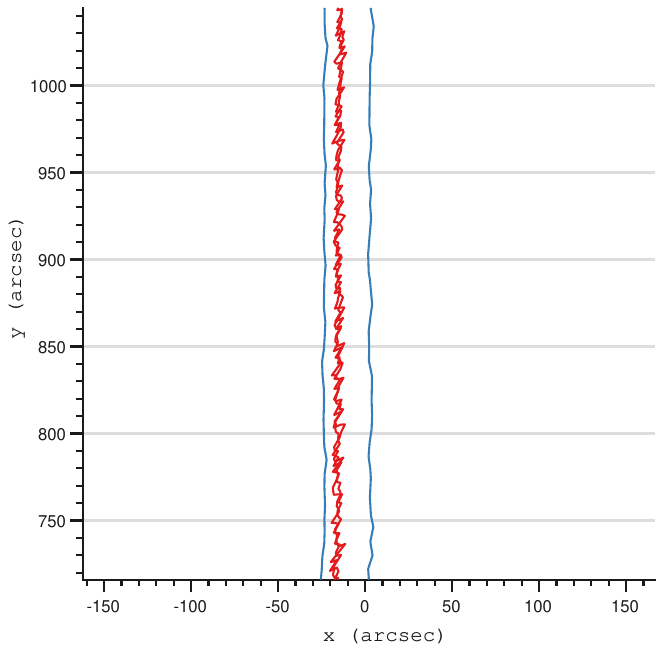}
     \includegraphics[width=0.23\textwidth]{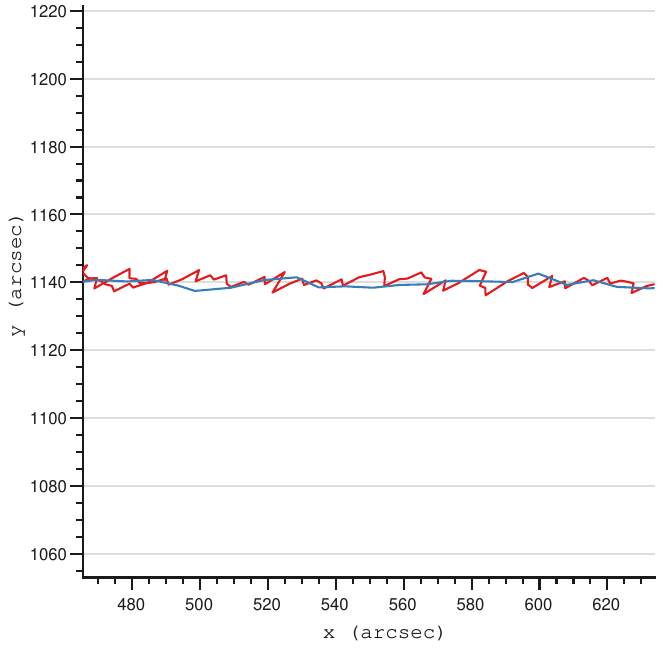}
     \includegraphics[width=0.23\textwidth]{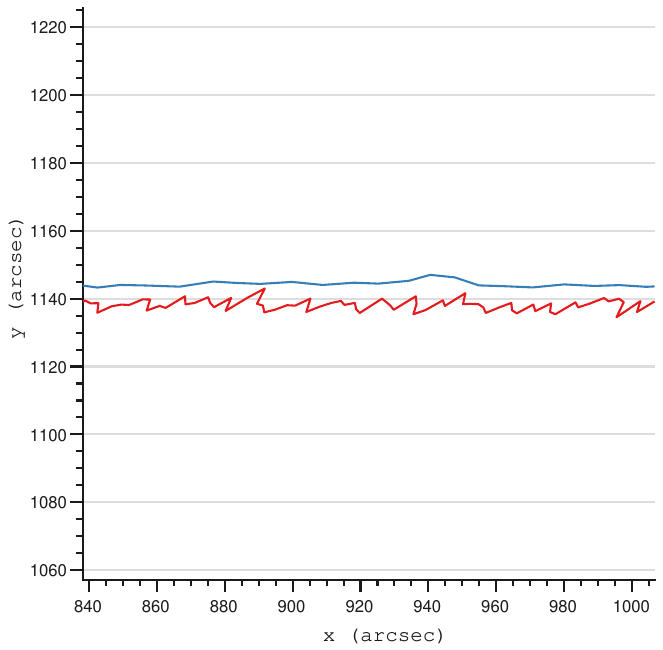}
     } \\
     \subfloat[][\texttt{circle} trajectory]{
     \includegraphics[width=0.28\textwidth]{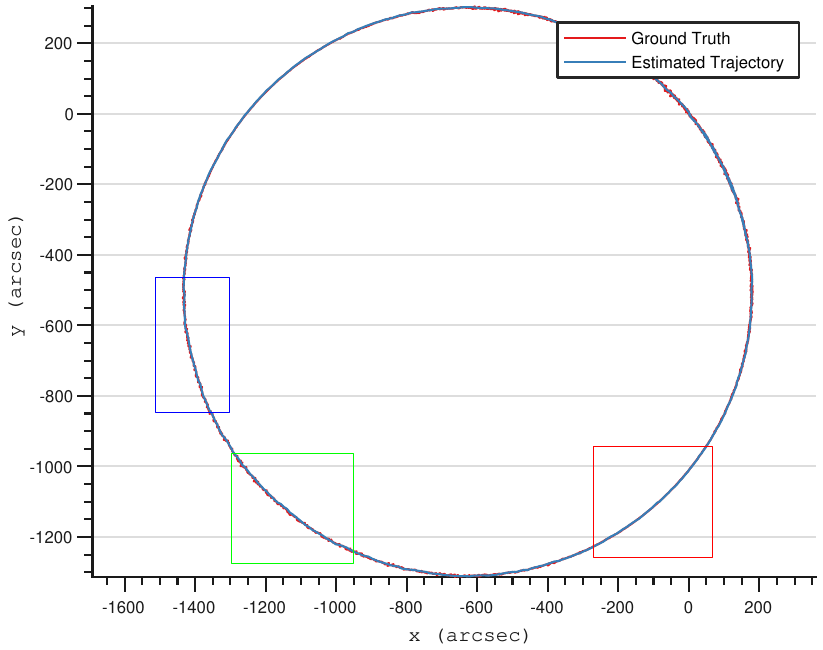}
     \includegraphics[width=0.23\textwidth]{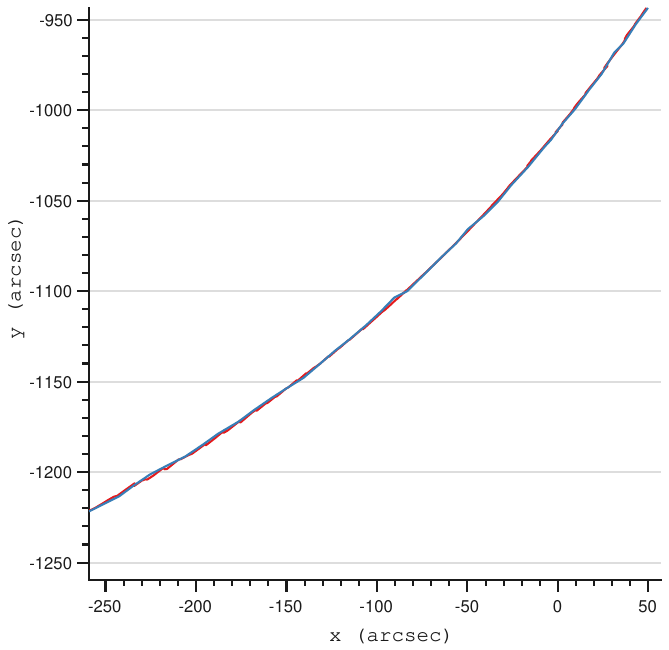}
     \includegraphics[width=0.23\textwidth]{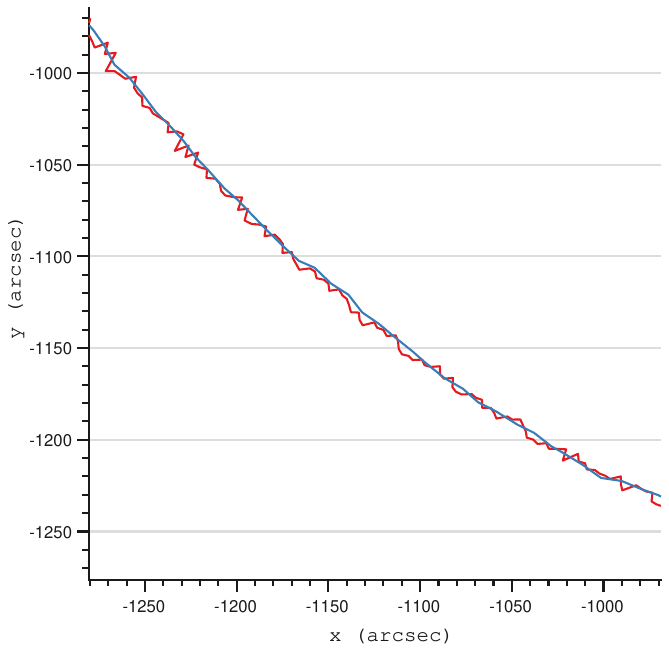}
     \includegraphics[width=0.23\textwidth]{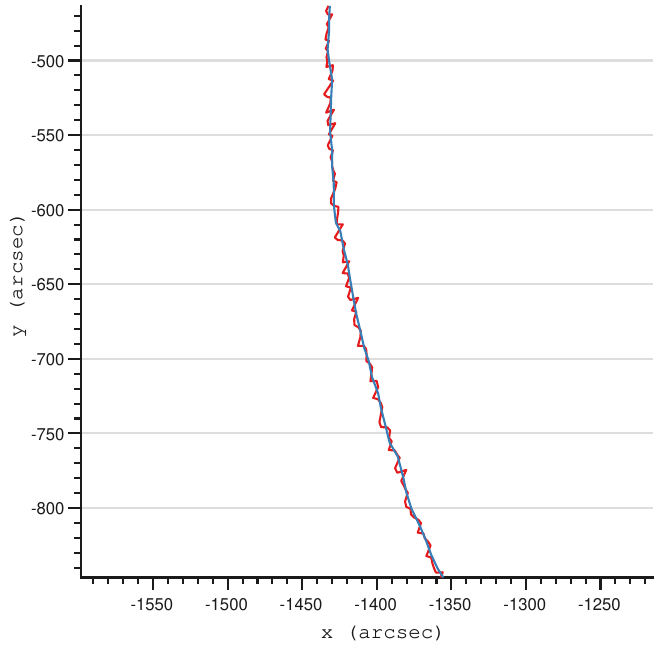}
     } \\
     \caption{Demonstrative results for various test trajectories at $\sigma = 10^{-8}$ degrees. First column demonstrates the ground truth trajectory in red and the estimated trajectory in blue. The subsequent columns show the insets marked in red, green and blue respectively.}
     \label{fig:results-tracking}
\end{figure*}

To quantify the open loop tracking accuracy, the estimated trajectory is aligned against the ground truth and the Root Mean Squared Error (RMSE) metric is reported for various noise levels $\sigma_N$. 
\begin{equation}
    e_{\sigma_N} = \sum_{t=0}^{N}|| \mathbf{p}_t^{\sigma} - \mathbf{g}_t^{\sigma} ||_{_2}
\end{equation}
where $\mathbf{p}_t^{\sigma_N}$ is the computed position against the ground truth position $\mathbf{g}_t^{\sigma_N}$ at time $t$ with a noise level $\sigma_N$.
The computed metric $e_{\sigma_N}$ provides an estimate of the average pointing discrepancy in the system at a particular noise level. 
Tracking accuracy for the \texttt{linear}, \texttt{square} and \texttt{circle} trajectories for $\sigma = 10^{-8}$ degrees is reported in the Fig. \ref{fig:results-tracking}. Three inset marked in the first column provide details of the estimated position for a short time period in the other three columns. Additionally, results for various trajectories is summarised in Fig. \ref{fig:rmse} which reports the RMSE tracking error across various noise levels for the \texttt{linear}, \texttt{circle} and \texttt{square} trajectories as well the high frequency tracking task described in Sec. \ref{sec:hf}.

% We also look at the tracking performance across time for the square trajectory. As before, we plot the x and y positions (pixels) across time. The zoomed in view provide the estimated and the ground truth trajectory between the 3- and 4-seconds mark. Under low noise situation, the system can track the motion precisely, keeping the RMSE to below 1 pixel which well within the desired level of accuracy for the system.  

10 arcsec line represents the precision that the system is aiming for. It can be seen that the tracking error remains below the 10 arcsec cut-off for noise levels up to $\sigma = 10^{-7}$ degrees. Tracking accuracy degrades with increased noise. As expected, high frequency noise is more difficult to instantaneously estimate. On-board a typical small satellite, the noise is low frequency and falls within the operational range of the tracker (below $\sigma_N = 10^{-7}$ degrees). Results show that the system can reliably estimate the motion of the sensor within the required 10-arcsec level of accuracy. Tracking high-frequency perturbations leads to greater errors compared to the tracking the smoothed signal via the Kalman filter  for the \texttt{linear} trajectory. 

\subsection{Closed Loop Stabilisation accuracy}
Having a good grasp of the tracking capabilities of the proposed system, we demonstrate its effectiveness in providing ultra-fine pointing stabilisation. This requires the translation of positional estimation into stabilisation command for the piezoelectric stage. 

\begin{figure}
     \centering
     \subfloat[][\texttt{linear} trajectory]{
     \includegraphics[width=0.40\textwidth]{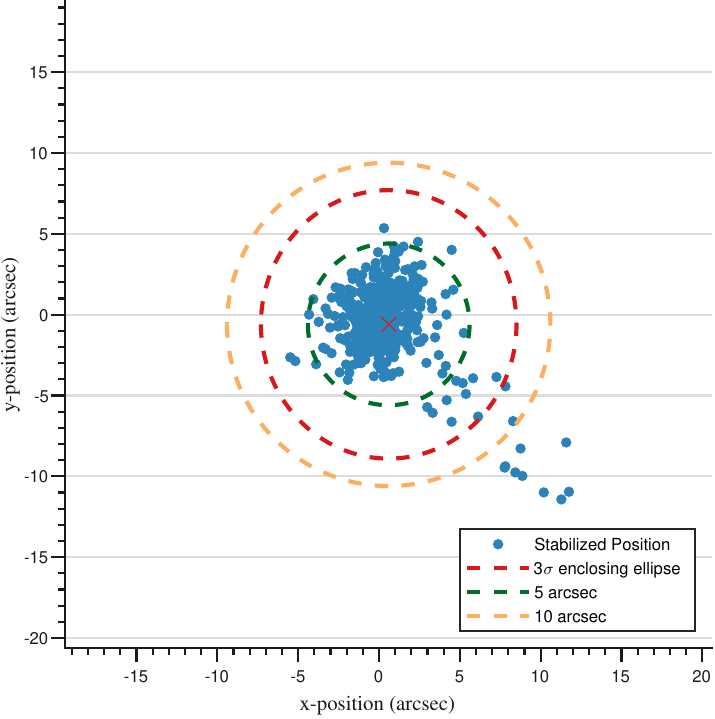}
     \includegraphics[width=0.47\textwidth]{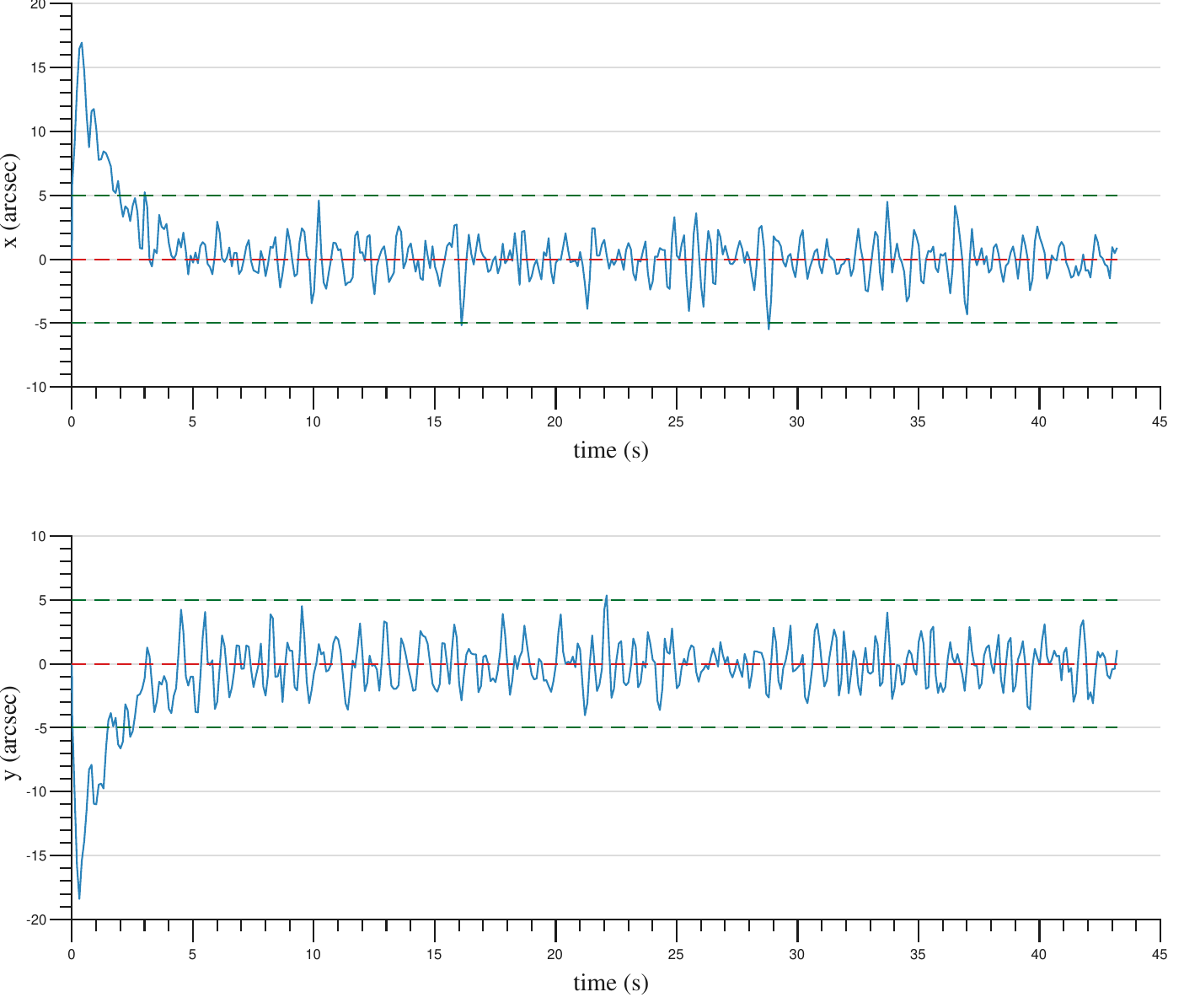}
     } \\
     \subfloat[][\texttt{square} trajectory]{
     \includegraphics[width=0.40\textwidth]{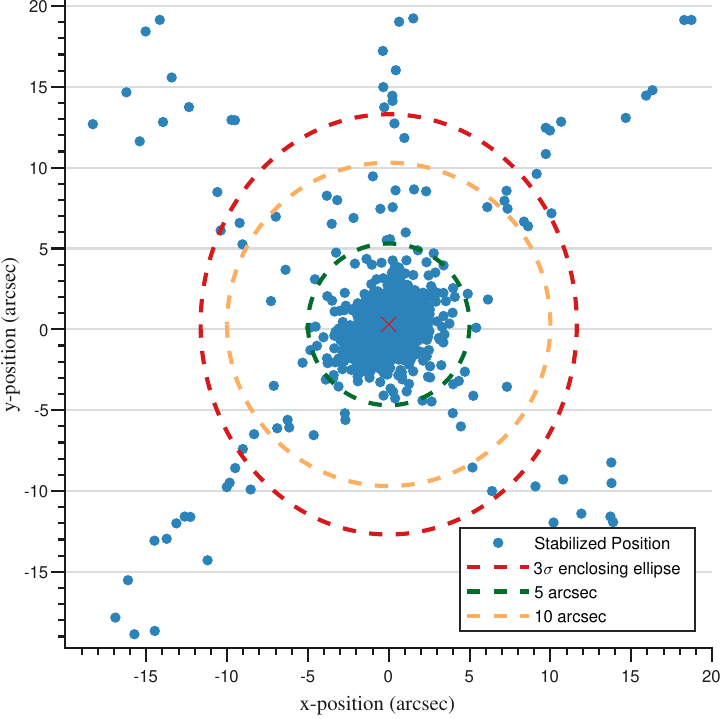}
     \includegraphics[width=0.47\textwidth]{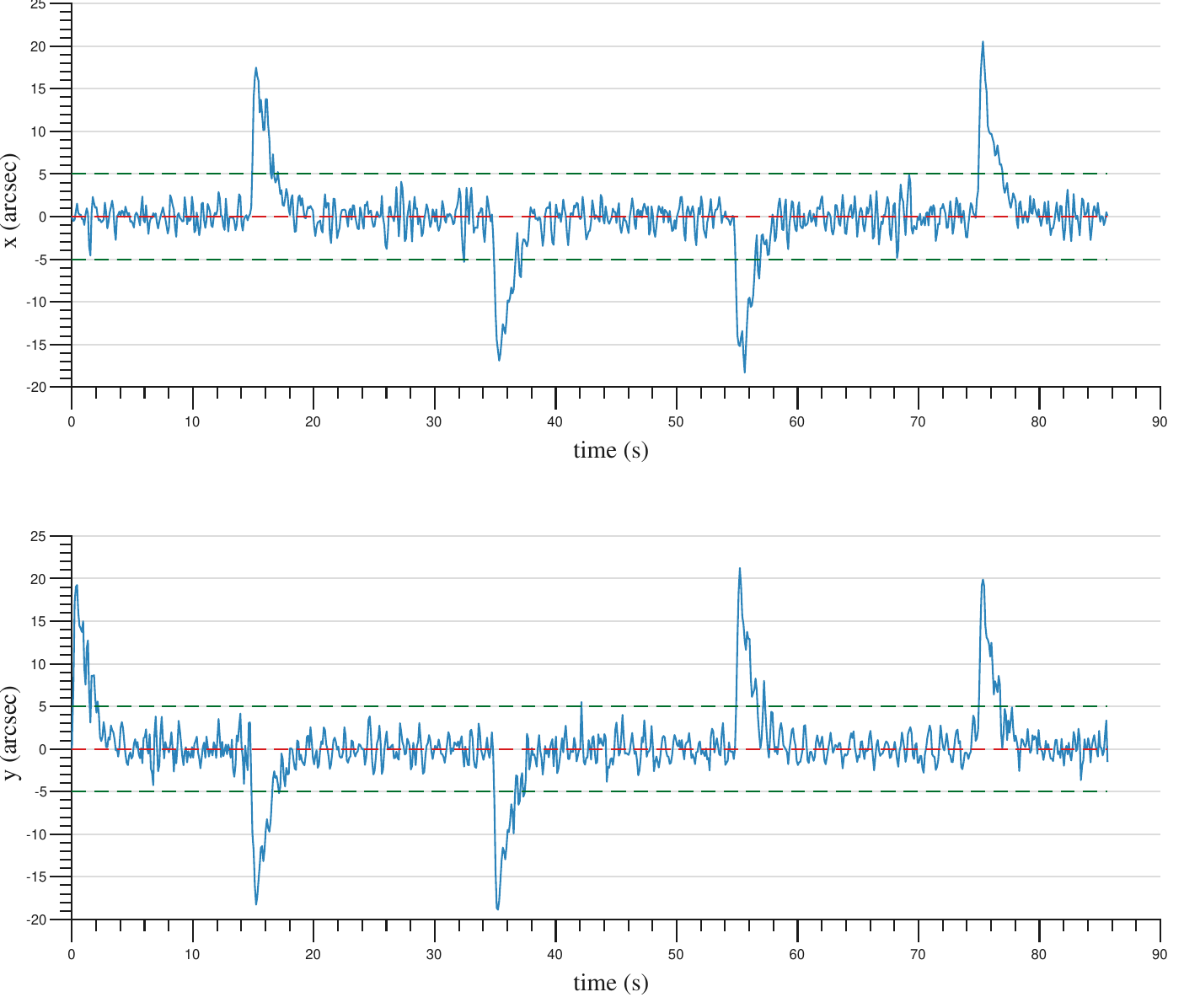}
          } \\
     \subfloat[][\texttt{circle} trajectory]{
     \includegraphics[width=0.40\textwidth]{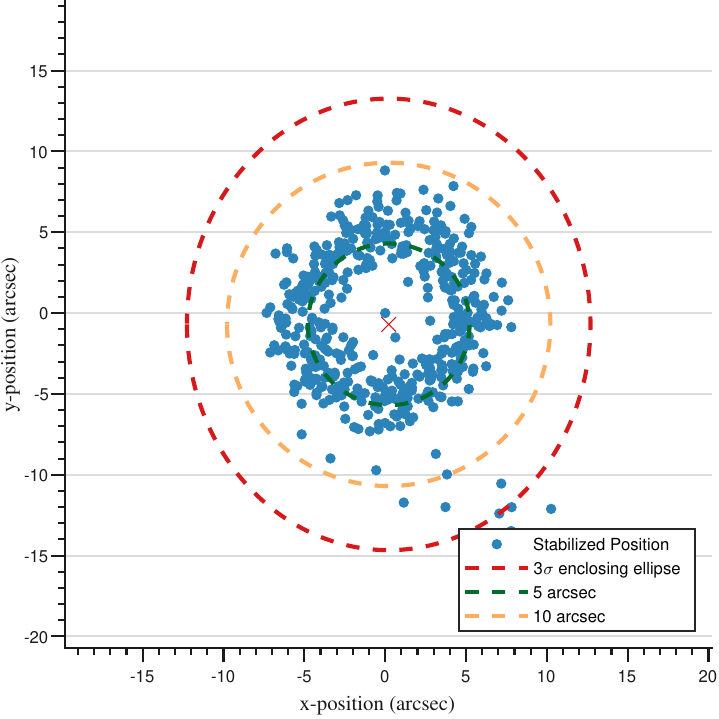}
     \includegraphics[width=0.47\textwidth]{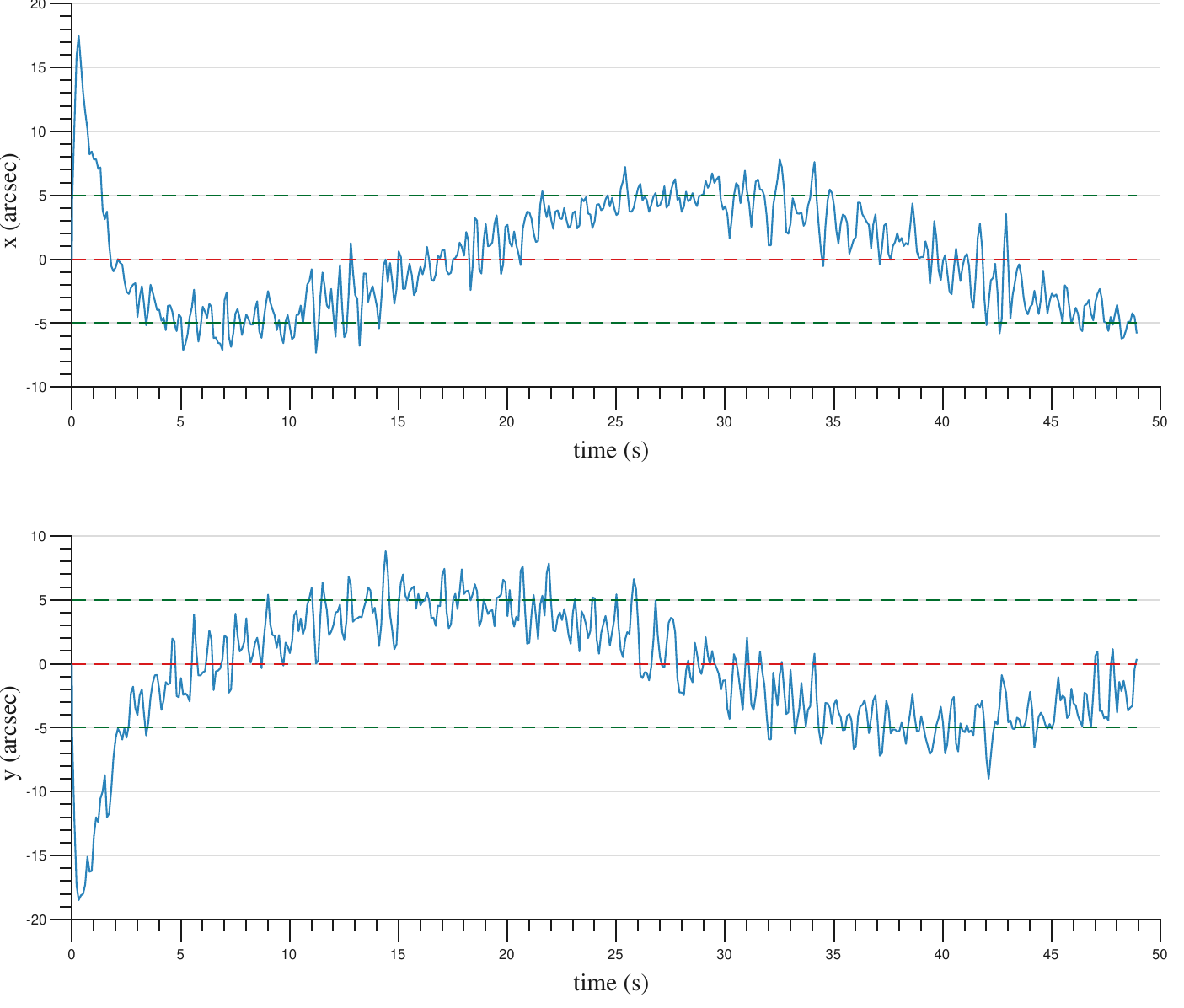}
     } \\
     \caption{Stabilisation results at $\sigma_N = 10^{-8}$ degrees: Left: Error Ellipse (red) should ideally lie inside the yellow circle marking a radius of 10 arcseconds. Right: Evolution of error over time during stabilisation. Once stabilised, the error should remain within remain within the bound shown.}
     \label{fig:results-stabilisation}
\end{figure}

To quantify the stabilisation performance of the system, we report the spread of the deviation from the required pointing direction over time. The smaller the spread of the points across the mean, the better the pointing position is maintained over time. We report this spread, effectively the $3\sigma$ contours of the estimated Gaussian fit to the deviation from the required pointing position, for various trajectories in Fig. \ref{fig:results-stabilisation}. 
As before, we aim for most of the stabilised positions to fall within 10 arcseconds radius of the required pointing direction on the event sensor, as indicated by the yellow circle in the Fig. \ref{fig:results-stabilisation}(left). On the right, the  evolution of stabilisation across time is shown. When stabilisation is requested, the system is driven to the required position by the PID controller over time. This leads to a reduction in the pointing error seen at the beginning. The system then actively computes and corrects for any deviations from the pointing direction. This stabilised trajectory with some perturbations can be seen in the plots. High magnitude deviations, such as those
seen in the \texttt{square} trajectory, when sudden changes in position occur at the corners, leads to deviation in the pointing direction which are subsequently corrected, and  stable pointing is maintained. It should be noted the under the simulated noise the $1\sigma$ performance remains well below 10 arcseconds in Table \ref{tab:stabilisation-performance}. 

\begin{table}[!h]
    \centering
  \begin{tabular}{cc | c|c|c}
    & & \multicolumn{3}{c}{noise level (degrees)} \\\hline
    \multicolumn{2}{c|}{trajectory} & $10^{-9}$ & $10^{-8}$ & $10^{-7}$  \\ \hline
    %------
    \multirow{2}{*}{\texttt{linear}}
        & $\sigma_x$ & 3.36 & 2.61 & 3.62 \\ \cline{2-5}
        & $\sigma_y$ & 2.61 & 2.73 & 3.70\\ \hline
    \multirow{2}{*}{\texttt{square}}
        & $\sigma_x$ & 3.28 & 3.87 & 4.59\\ \cline{2-5}
        & $\sigma_y$ & 3.74 & 4.29 & 5.22\\ \hline
    \multirow{2}{*}{\texttt{circle}}
        & $\sigma_x$ & 7.79 & 3.74 & 4.71 \\ \cline{2-5}
        & $\sigma_y$ & 4.21 & 4.63 & 5.01
    
\end{tabular}
    \caption{Stabilisation performance for various noise level for the simulated trajectories. Results report 1$\sigma$ deviation in arcseconds.}
    \label{tab:stabilisation-performance}
\end{table}

% \begin{figure}
%      \centering
%      \subfloat[][\texttt{linear} trajectory]{
%      \includegraphics[width=0.28\textwidth]{figs/results/linear_main_8.png}
%      \includegraphics[width=0.23\textwidth]{figs/results/linear_main_8_1.png}
%      \includegraphics[width=0.23\textwidth]{figs/results/linear_main_8_2.png}
%      \includegraphics[width=0.23\textwidth]{figs/results/linear_main_8_3.png}
%      } \\
%      \subfloat[][\texttt{square} trajectory]{
%      \includegraphics[width=0.28\textwidth]{figs/results/sqaure_main_8.png}
%      \includegraphics[width=0.23\textwidth]{figs/results/sqaure_main_8_1.png}
%      \includegraphics[width=0.23\textwidth]{figs/results/sqaure_main_8_2.png}
%      \includegraphics[width=0.23\textwidth]{figs/results/sqaure_main_8_3.png}
%      } \\
%      \subfloat[][\texttt{circle} trajectory]{
%      \includegraphics[width=0.28\textwidth]{figs/results/circle_main_8.png}
%      \includegraphics[width=0.23\textwidth]{figs/results/circle_main_8_1.png}
%      \includegraphics[width=0.23\textwidth]{figs/results/circle_main_8_2.png}
%      \includegraphics[width=0.23\textwidth]{figs/results/circle_main_8_3.png}
%      } \\
%      \caption{Demonstrative results for various test trajectories at $\sigma = 10^{-8}$ degrees. The first column demonstrates the ground truth trajectory in red and the estimated trajectory in blue. The subsequent column show the regions marked in red, green and blue respectively.}
%      \label{fig:results-tracking}
% \end{figure}

\subsection{Computational Time}
We run a benchmarking sequence 35 seconds long to benchmark the time taken by various component of the pipeline on different machines for the open loop star tracking task. The same code in run on two different machines: \texttt{desktop} representing a modern desktop machine equipped with an Intel Core i7-8700 CPU and Nvidia Titan X GPU and \texttt{Jetson} which is the Jetson Nano 2 with an onboard GPU. The code utilises GPUs for noise suppression (referred to as Median filtering in the Tab. \ref{tab:time}), while the rest of the code runs on the CPU. 
The \texttt{desktop} represents a reference value which can be treated as the upper bound for the satellite case.

% For each of the machines, we evaluate two configurations: \texttt{vis} representing the full system where in addition to open loop tracking, both the input and output of the system are visualised to be seen by the human operator and \texttt{no\_vis} where the algorithm run in a headless manner and the outputs are written to disk at the end of execution. This is the expected on-board behaviour.
%
Table \ref{tab:time} show that our proposal can run at nearly 50Hz, signifying the highest update rate at which the piezoelectric stage can be driven on-board. This update rate is faster than what would be available using a conventional sensing modality such as a CMOS/CCD sensor.

\begin{table}[!h]
    \centering
\begin{tabular}{c|c|c|c}
%Machine & mode & System time (Frequency) & Star Detection Time   (\%) & Median Filtering Time (GPU) (\%) \\ \hline
Machine & System time (Frequency) & Star Detection Time   (\%) & Median Filtering Time (GPU) (\%) \\ \hline
desktop  & 3.26 ms (306Hz) & 1.21 ms (37.11\%) & 1.77 ms (54.29\%) \\ \hline
%desktop & \texttt{vis} & 6.26 ms (160Hz) & 1.20 ms (19.16\%) & 1.70 ms (27.15\%) \\ \hline
Jetson  & 20.39 ms (49Hz) & 4.39 ms (21.53\%) & 14.58 ms (71.50\%) \\ 
%Jetson  & \texttt{vis} & 39.15 ms (25Hz) & 3.49 ms (08.91\%) & 12.00 ms (30.65\%) \\
    \end{tabular}
\caption{Processing Times on various machines}
\label{tab:time}

\end{table}

% In addition, Table \ref{tab:time} also provides information about two of the main compute heavy tasks in the pipeline and the percentage of time take by them in each configuration. The interesting case to be noticed is that for the \texttt{no\_viz} setting for Jetson, the noise suppression module (Median Filtering) takes up about 70\% of the compute resources and thus limits the operational frequency of the system to about 50 Hz. On the contrary, on a more powerful machine the same code can run at about 300 Hz. To further improve the operation frequency of the system on the Jetson, a simpler but effective noise suppression mechanism would need to be investigated to allow the computational resources to be better allocated to other parts of the pipeline. 

\section{Conclusion}
As the requirement for accurate pointing increases on board commercial missions, new sensing modalities need to be incorporated into payloads to enable high frequency high accuracy pointing. In this work, we have presented a novel payload using a combination of a event sensor and piezoelectric motion stage to achieve ultra-fine attitude estimation and ultra-fine pointing stabilisation. In contrast to the ADCS which achieves pointing stabilisation by manoeuvring the whole satellite body, our proposal offers additional pointing corrections by separately controlling the imaging sensor, alleviating the residual pointing errors in the ADCS.
We have demonstrated the feasibility of COTS hardware and developed algorithms that can run efficiently onboard at a much higher frequency than commercially available star tracking based pointing mechanism.  With detailed experiments, we have reported results for the open loop star tracking as well as the closed loop stabilisation tasks. We have shown that the system achieves attitude estimation and stabilisation that can complement existing ADCS solutions.
\section{Acknowledgements}
This work has been supported by the SmartSat CRC, whose activities are funded by the Australian Government’s CRC Program. Tat-Jun Chin is the SmartSat CRC Professorial Chair of Sentient Satellites.
\clearpage

\bibliographystyle{plain}
\bibliography{references}

\begin{thebibliography}{10}

\bibitem{afshar2020event}
Saeed Afshar, Andrew~Peter Nicholson, Andre Van~Schaik, and Gregory Cohen.
\newblock Event-based object detection and tracking for space situational
  awareness.
\newblock {\em IEEE Sensors Journal}, 20(24):15117--15132, 2020.

\bibitem{PIDController}
Stuart Bennett.
\newblock Development of the pid controller.
\newblock {\em IEEE Control Systems Magazine}, 13(6):58--62, 1993.

\bibitem{past-present-slam}
Cesar Cadena, Luca Carlone, Henry Carrillo, Yasir Latif, Davide Scaramuzza,
  Jos{\'e} Neira, Ian Reid, and John~J Leonard.
\newblock Past, present, and future of simultaneous localization and mapping:
  Toward the robust-perception age.
\newblock {\em IEEE Transactions on robotics}, 32(6):1309--1332, 2016.

\bibitem{cohen2019event}
Gregory Cohen, Saeed Afshar, Brittany Morreale, Travis Bessell, Andrew Wabnitz,
  Mark Rutten, and Andr{\'e} van Schaik.
\newblock Event-based sensing for space situational awareness.
\newblock {\em The Journal of the Astronautical Sciences}, 66:125--141, 2019.

\bibitem{RANSAC}
Martin~A Fischler and Robert~C Bolles.
\newblock Random sample consensus: a paradigm for model fitting with
  applications to image analysis and automated cartography.
\newblock {\em Communications of the ACM}, 24(6):381--395, 1981.

\bibitem{gallego2020event}
Guillermo Gallego, Tobi Delbr{\"u}ck, Garrick Orchard, Chiara Bartolozzi, Brian
  Taba, Andrea Censi, Stefan Leutenegger, Andrew~J Davison, J{\"o}rg Conradt,
  Kostas Daniilidis, et~al.
\newblock Event-based vision: A survey.
\newblock {\em IEEE transactions on pattern analysis and machine intelligence},
  44(1):154--180, 2020.

\bibitem{tycho2}
Erik H{\o}g.
\newblock Tycho star catalogs: The 2.5 million brightest stars.
\newblock In {\em Encyclopedia of Astronomy \& Astrophysics}, pages 1--3. CRC
  Press, 2001.

\bibitem{jawaid2023towards}
Mohsi Jawaid, Ethan Elms, Yasir Latif, and Tat-Jun Chin.
\newblock Towards bridging the space domain gap for satellite pose estimation
  using event sensing.
\newblock In {\em 2023 IEEE International Conference on Robotics and Automation
  (ICRA)}, pages 11866--11873. IEEE, 2023.

\bibitem{KF}
Richard~J Meinhold and Nozer~D Singpurwalla.
\newblock Understanding the {K}alman filter.
\newblock {\em The American Statistician}, 37(2):123--127, 1983.

\bibitem{AKF}
Yonhon Ng, Yasir Latif, Tat-Jun Chin, and Robert Mahony.
\newblock Asynchronous kalman filter for event-based star tracking.
\newblock In {\em Computer Vision – ECCV 2022 Workshops: Tel Aviv, Israel,
  October 23–27, 2022, Proceedings, Part I}, page 66–79, Berlin,
  Heidelberg, 2023. Springer-Verlag.

\bibitem{papotti2021star}
Gabor Papotti.
\newblock {\em A Star Tracker based Attitude Determination System}.
\newblock PhD thesis, ResearchSpace@ Auckland, 2021.

\bibitem{pong2018orbit}
Christopher Pong.
\newblock On-orbit performance \& operation of the attitude \& pointing control
  subsystems on {ASTERIA}.
\newblock {\em Proceedings of the Small Satellite Conference}, 2018.

\bibitem{pong2010achieving}
Christopher~M Pong, Sungyung Lim, Matthew~W Smith, David~W Miller, Jesus~S
  Villase{\~n}or, and Sara Seager.
\newblock Achieving high-precision pointing on exoplanetsat: initial
  feasibility analysis.
\newblock In {\em Space Telescopes and Instrumentation 2010: Optical, Infrared,
  and Millimeter Wave}, volume 7731, pages 620--635. SPIE, 2010.

\end{thebibliography}

\end{document}